\documentclass{aastex701}

\usepackage{amssymb}
\usepackage{amsmath}
\usepackage{verbatim}
\usepackage{mathrsfs}
\usepackage{amsfonts}
\usepackage{amsfonts} 
\usepackage{latexsym}
\usepackage{epsfig}
\usepackage{color}
\usepackage{graphicx}
\usepackage{envmath}
\usepackage{natbib}
\usepackage{tikz}
\usepackage{soul}
\usepackage[normalem]{ulem}
\usepackage{bigints}
\usepackage{makecell}
\usepackage{multirow}
\usepackage{subfig}

\usepackage[T1]{fontenc}
\usepackage{textcomp}

\usepackage{siunitx,tabularx,ragged2e}
\usepackage{array}
\usepackage{hyperref}

\usepackage{booktabs}

\begin{document}

%\title{Forecasts of the combined analysis of GRBs and CMB on the wCDM model}
\title{Gamma-Ray Bursts as an Independent High-Redshift Probe of Dark Energy}

\author[orcid=0000-0003-4442-8546]{Maria Giovanna Dainotti}
\email[show]{maria.dainotti@nao.ac.jp}
\affiliation{National Astronomical Observatory of Japan, 2 Chome-21-1 Osawa, Mitaka, Tokyo 181-8588, Japan}
\affiliation{The Graduate University for Advanced Studies, SOKENDAI, Shonankokusaimura, Hayama, Miura District, Kanagawa 240-0193, Japan}
\affiliation{Space Science Institute, 4765 Walnut St, Suite B, 80301 Boulder, CO, USA}
\affiliation{Department of Physics and Astrophysics, University of Las Vegas, NV 89154, USA}

\author[orcid=0000-0003-1943-010X]{Aleksander Łukasz Lenart} 
\affiliation{Jagiellonian University, Doctoral School of Exact and Natural Sciences, Kraków, Poland}
\affiliation{Astronomical Observatory of Jagiellonian University in Kraków, Orla 171, 30-244 Kraków}
\email[show]{aleksander.lenart@doctoral.uj.edu.pl}

\author[orcid=0000-0001-5083-6461]{Biagio De Simone}
\affiliation{Dipartimento di Fisica ``E.R. Caianiello”, Università di Salerno, Via Giovanni Paolo II, 132, Fisciano, Salerno, 84084, Italy}
\affiliation{INFN Gruppo Collegato di Salerno - Sezione di Napoli. c/o Dipartimento di Fisica ”E.R. Caianiello”, Università di Salerno, Via Giovanni Paolo II, 132, Fisciano, Salerno, 84084, Italy}
\email{biagio.des94@gmail.com}

\author[orcid=0000-0002-4012-9285]{William Giar\`{e}}
\email{giare@hawaii.edu}
\affiliation{Department of Physics and Astronomy, University of Hawai‘i, Honolulu, HI 96822, USA}

\author[0000-0001-8408-6961]{Eleonora Di Valentino}
\email{e.divalentino@sheffield.ac.uk}
\affiliation{School of Mathematics and Statistics, University of Sheffield, Hounsfield Road, Sheffield S3 7RH, United Kingdom}

\author[0000-0002-8028-0991]{Dieter H. Hartmann}\email{hdieter@clemson.edu}
\affiliation{Department of Physics and Astronomy, Clemson University, Clemson, SC 29634, USA}

\author[0000-0002-0173-6453]{Nissim Fraija}
\affiliation{Instituto de Astronom\'ia, Universidad Nacional Aut\'{o}noma de M\'{e}xico, A. P. 70-264, Cd. Universitaria, Ciudad de M\'{e}xico, M\'{e}xico}
\email{nifraija@astro.unam.mx}

\author[0000-0002-2707-7548]{Kazunari Iwasaki}
\affiliation{National Astronomical Observatory of Japan, 2 Chome-21-1 Osawa, Mitaka, Tokyo 181-8588, Japan}
\email{kazunari.iwasaki@nao.ac.jp}

\author[0000-0001-7574-2330]{Gaetano Lambiase}
\affiliation{Dipartimento di Fisica ``E.R. Caianiello”, Università di Salerno, Via Giovanni Paolo II, 132, Fisciano, Salerno, 84084, Italy}
\affiliation{INFN Gruppo Collegato di Salerno - Sezione di Napoli. c/o Dipartimento di Fisica ”E.R. Caianiello”, Università di Salerno, Via Giovanni Paolo II, 132, Fisciano, Salerno, 84084, Italy}
\email{glambiase@unisa.it}
%% Use the \collaboration command to identify collaborations. This command
%% takes an optional argument that is either a number or the word "all"
%% which tells the compiler how many of the authors above the command to
%% show. For example "\collaboration[all]{(DELVE Collaboration)}" wil include
%% all the authors above this command.
%%
%% Mark off the abstract in the ``abstract'' environment. 

\begin{abstract}
Testing the $\Lambda$CDM model requires cosmological probes spanning the wide redshift interval between Type Ia Supernovae (SNe Ia, $z\lesssim2.9$) and the Cosmic Microwave Background (CMB, $z\approx1100$). Gamma-Ray Bursts (GRBs), observed up to redshift $z=9.2$, offer the opportunity to explore this regime. Here, we investigate how many GRBs are needed to become a useful cosmological probe capable of independently testing deviations from $\Lambda$CDM suggested by the recent DESI BAO observations.
We develop forecasts based on the two-dimensional X-ray and optical Dainotti relations, between the luminosity at the end of the plateau phase and its rest-frame duration. Using simulated GRB samples constructed from the observed population, we evaluate the constraining power of GRBs on cosmological parameters within the $w$CDM and $w_0w_a$CDM models, both independently and in combination with CMB observations.
Our results show that GRB samples containing several tens to hundreds of well-characterized plateau can already approach the precision currently achieved by CMB measurements on the Dark Energy (DE) equation-of-state parameter $w$. Particularly, a sample of $\sim66$ optical GRBs can reach a precision $\sigma_w \approx 0.47$, comparable to that obtained from Planck within the $w$CDM framework. Such sample sizes are already attainable through Machine Learning techniques that double the number of GRBs using inferred redshifts.
These forecasts indicate that future GRB observations, when combined with next-generation transient missions and improved statistical techniques, will provide an independent high-redshift probe of cosmic expansion and will play an important role in testing the robustness of potential Dynamical DE signals suggested by other cosmological datasets. 

\end{abstract}

%% Keywords should appear after the \end{abstract} command. 
%% The AAS Journals now uses Unified Astronomy Thesaurus (UAT) concepts:
%% https://astrothesaurus.org
%% You will be asked to selected these concepts during the submission process
%% but this old "keyword" functionality is maintained in case authors want
%% to include these concepts in their preprints.
%%
%% You can use the \uat command to link your UAT concepts back its source.

%% From the front matter, we move on to the body of the paper.
%% Sections are demarcated by \section and \subsection, respectively.
%% Observe the use of the LaTeX \label
%% command after the \subsection to give a symbolic KEY to the
%% subsection for cross-referencing in a \ref command.
%% You can use LaTeX's \ref and \label commands to keep track of
%% cross-references to sections, equations, tables, and figures.
%% That way, if you change the order of any elements, LaTeX will
%% automatically renumber them.
%=============================================
%Introduction could highlight novelty more strongly
% Currently the novelty is implicit.
% You should explicitly say:
% Possible sentence:
% This work differs from previous GRB cosmology analyses by forecasting the precision achievable with next-generation GRB samples and by quantifying the number of sources required to reach Planck-level constraints.

\section{Introduction}\label{introduction}
Recently, increasingly precise measurements of cosmological parameters have uncovered significant inconsistencies in the widely accepted cosmological model known as the flat $\Lambda$ Cold Dark Matter ($\Lambda$CDM) model. This framework describes the Universe by incorporating Cold Dark Matter (CDM) and Dark Energy components, with Dark Energy represented as a cosmological constant ($\Lambda$), which accounts for the current accelerated expansion of the Universe~\citep{riess1998, perlmutter1999}. 
The flat $\Lambda$CDM model successfully fits observations such as the Cosmic Microwave Background (CMB) radiation~\citep{Planck:2018vyg}, the Baryon Acoustic Oscillations (BAO; \citealt{eboss2021}), and the accelerated expansion of the Universe inferred from Type Ia Supernovae (SNe Ia). Nevertheless, it presents well-known theoretical shortcomings. One example is the cosmological constant problem~\citep{1989RvMP...61....1W}, which reflects the discrepancy between the expected and the observed values of $\Omega_{\Lambda}$, as well as the unresolved question of the physical origin of Dark Energy. Another issue is the fine-tuning (or coincidence) problem, which arises from the fact that the current values of the matter density ($\Omega_{M}$) and the Dark Energy density ($\Omega_{\Lambda}$) are of the same order of magnitude, despite their different evolution in time.
To further test whether the flat $\Lambda$CDM model provides an adequate description of the Universe, reliable cosmological probes are required at redshifts between $z = 2.9$ \citep{Rodney} (the maximum redshift of SNe Ia so far observed for cosmological studies, \citep{Dainotti2025JHEAp}) and $z = 1100$. In this paper, we employ Gamma-Ray Bursts (GRBs) as a potential probe to tackle this issue. 

GRBs represent one of the few astrophysical probes capable of exploring the expansion history of the Universe at very high redshift. While SNe Ia provide precise cosmological measurements up to $z \approx 2.9$~\citep{2025A&A...701A..70V}, GRBs have been detected up to $z \approx 9.4$~\citep{2011ApJ...736....7C}, and may extend even further with future missions ($z \approx 15$, \citealt{Lamb2000,Lamb_2003}). This makes GRBs particularly valuable for bridging the redshift gap between SNe Ia and the CMB. 
In recent years, correlations involving the plateau phase of GRB afterglows, most notably the luminosity-time relation (the Dainotti relation), have emerged as promising tools for standardizing GRB energetics~\citep{Cardone2009,2010MNRAS.408.1181C,Dainotti2013b,Postnikov2014,Dainotti2023c}. These relations exhibit relatively small intrinsic scatter and have been shown to remain robust against selection effects when appropriate statistical methods are applied.
Afterglow correlations have the advantage of exhibiting less variability than correlations derived from the prompt emission phase.
As a result, GRBs can be used to construct high-redshift Hubble diagrams that provide cosmological constraints independent of traditional probes. In this context, GRBs offer a complementary avenue for testing extensions of the $\Lambda$CDM model, including the $w$CDM and Dynamical Dark Energy (DDE) scenarios, and may play an increasingly important role in cosmology as future missions increase the available GRB sample.

An important advantage of GRBs as cosmological probes is that they are affected by systematics that are independent of those associated with Type Ia Supernovae. The standardization of SNe Ia relies on empirical relations between light-curve shape, colour, and luminosity, and is subject to systematic uncertainties related to dust extinction, host-galaxy properties, population evolution, and photometric calibration. In contrast, GRB cosmology relies on correlations between physical properties of relativistic jets, the inner engine, and their afterglow emission, such as the luminosity-time relation associated with the plateau phase. Consequently, GRBs provide an independent probe of cosmic expansion that can be used to further constrain or verify cosmological constraints derived from other distance indicators. The use of multiple independent probes is particularly important in the current context of tensions in cosmological parameters, such as the Hubble constant discrepancy~\citep{Verde:2019ivm,DiValentino:2020zio,Dainotti_2021,DiValentino:2021izs,Perivolaropoulos:2021jda,Abdalla:2022yfr,galaxies10010024,DiValentino:2022fjm,Kamionkowski:2022pkx,Giare:2023xoc,2024arXiv241105744D,Verde:2023lmm,2024arXiv241105744D,DiValentino:2024yew,2024PDU....4401486M,Dainotti2025JHEAp, CosmoVerseNetwork:2025alb}, since independent measurements can help determine whether the observed discrepancies arise from new physics or from unaccounted systematic effects.
Indeed, some recent studies have shown redshift-dependent trends in the Hubble constant parameter when SNe Ia and BAO are considered ~\citep{Dainotti_2021,galaxies10010024,Dainotti2025JHEAp}. The interpretation of these results further highlights the importance of independent cosmological probes, such as GRBs \citep{ 2022PASJDainotti,Dainotti2023alternative}, Quasars (QSO, \citealt{2022ApJ...931..106D,2023ApJS..264...46L,2023ApJ...950...45D,2024Galax..12....4D}), Gravitational Waves (GW, \citealt{2020ApJS..250....6W,2021ApJ...921..149N}), and the combination of Cosmic Chronometers (CC) and BAOs into a unique set called "Hubble data" \citep{2024MNRAS.533.2232Y}, capable of testing the expansion history across a wide range of redshifts.
%Indeed, in recent years, we have witnessed that there is an evolving trend of $H_0$ as a function of the redshift~\citep{Dainotti_2021,galaxies10010024}, which cannot actually be explained within the context of the $w$CDM and $w_0w_a$CDM models. This is the reason why this work is also preparatory to check the impact of the combination of Planck data and GRBs in terms of obtaining further constraints on $w$CDM model. 
%\textcolor{green}{[I'd move this sentence about H0 tension papers before the one that refers to CosmoVerse paper to make the periods better connected. In that case I'd add a small piece like "It is important to notice that in recent years we have witnessed..."]} \EDV{The Hubble tension has nothing to do with this paper. I kept the citations but removed the sentence.}

The standardization of GRB afterglows is supported by theoretical interpretations. Many authors have proposed theoretical explanations for the X-ray rest-frame end time of the plateau, $T_{a,X}^{*}$, and its corresponding luminosity, $L_{a,X}$, named Dainotti relation, built with Swift GRBs that exhibit a plateau phase and have a measured redshift ~\citep{Dainotti2008,Dainotti:2010ki,2011ApJ...730..135D,Dainotti2013a,2016ApJ...828...36D,Dainotti2020b}. This 2D correlation has been used to study and standardize GRB luminosities. This relation has also been extensively tested against selection biases using robust statistical methods such as the method introduced by ~\cite{Efron1992ApJ...399..345E}, see \citealt{Dainotti2013a,Dainotti2017a}. 
The discovery of the 2D Dainotti relation in X-rays marked the first time that an afterglow correlation was used as a cosmological tool. A detailed discussion of this topic is presented in ~\cite{Cardone2009,2010MNRAS.408.1181C,Dainotti2013b,Postnikov2014}. Although these works deal strictly with the Dainotti relation in X-rays, it has recently been found that a similar two-dimensional correlation also exists in optical wavelengths between the optical rest-frame end time, $T^{*}_{a,opt}$, and the optical luminosity at the end of the plateau, $L_{a,opt}$~\citep{Dainotti2020b,Levine2022}. This correlation has a slope similar to the X-ray Dainotti luminosity-time relation. 
For example,~\cite{2014MNRAS.443.1779R,stratta2018} interpreted it within a magnetar spin-down model \cite{2009ApJ...703.1696Z},~\cite{Lenart2025} presented a similar model based on a black hole central engine,~\cite{Cannizzo2009,Cannizzo2011} proposed an accretion-driven origin, while~\cite{Hascoet2014} analysed the possibility that the correlation arises from shocks interacting with the interstellar medium. 
There are many additional works analysing this correlation and investigating its physical origin~\citep{Zaninoni2011,Bernardini2012,Xu2012A&A...538A.134X,Mangano2012,Sultana_2012,Zaninoni2013,Margutti2013,vaneerten2014b,Bardho2015,Izzo2015,Kawakubo_2015,Si_2018,Zhao2019ApJ...883...97Z,Tang2019ApJS..245....1T,WangFeifei2020,Muccino2020,CaoShulei2022a,XuFan2021ApJ...920..135X,Levine2022,Shuang2022ApJ...924...69Y,Deng2023ApJ...943..126D,Tian_2023,Li_2023,Xu_2023,Deng2025}. 
These studies have motivated the use of this correlation as a cosmological tool in the literature~\citep{Cardone2009,Postnikov2014,Zitouni2016,Luongo2020,Cao2021,Luongo2021a,Luongo2021,Hu2021,Khadka2021,Muccino_2021,XuFan2021ApJ...920..135X,Wang2022ApJ...924...97W,CaoShulei2022,Dainotti2023c,Tian_2023,Li_2023,Xu_2023,Bargiacchi:2023jse,2023ApJ...951...63D,2024JCAP...08..015A,2024JHEAp..44..323F,LiJia2024,Alfano2024,SUDHARANI2024101522,Sethi2024}.

The 2D relation has been extended up to three dimensions with the addition of the peak prompt luminosity, $L_p$: the 3D Dainotti relation, known also as the \textit{fundamental plane relation}, has been confirmed both in X-ray and optical wavelengths~\citep{2016ApJ...825L..20D,2017ApJ...848...88D,2022ApJS..261...25D,DainottiVia2022MNRAS.514.1828D,Dainotti2023c}.
We here focus on the use of the 2D relation in optical and X-rays, which allows us to use GRBs as standard candles to constrain cosmological parameters, in analogy to what has been done with SNe Ia through the Phillips relation~\citep{Phillips1993}. Here, GRBs are used to investigate their constraining power on $w$ in the $w$CDM and $w_0w_a$CDM models, as well as in combination with CMB data to improve cosmological constraints. 
The $w$CDM model represents an extension of $\Lambda$CDM in which the Dark Energy Equation of State parameter $w$ differs from $-1$, but remains constant in time. Therefore, it does not capture the possible redshift evolution of DDE models. More general scenarios within DDE consider a time-evolving Equation of State. Recent analyses of the 2024 data release of the Dark Energy Spectroscopic Instrument (DESI, \citealt{DESI2025}) have highlighted a $3.2-3.4\sigma$ preference for such DDE models when DESI BAO measurements are combined with other cosmological probes, such as CMB and SNe Ia data~\citep{DES:2025sig,Hoyt:2026fve}.
The use of GRBs as cosmological probes to constrain the value of $w$ has been investigated by several authors.~\cite{Muccino_2021}, through the Combo~\citep{Izzo2015} relation, an extension of the 2D X-ray Dainotti by adding the spectral peak energy of the prompt emission ~\citep{Amati+02} for 174 GRBs, to constrain the value of $w$ at the low-redshift scale, showing that it is compatible with $w=-1$ at $z<1.2$, but starts deviating from it at higher redshifts.~\cite{Xie2025} provide a measurement of $w=-1.21^{+0.32}_{-0.67}$ using the Amati relation which relates the spectral peak energy to the isotropic equivalent radiated energy for 182 GRBs calibrated on the Pantheon+ SNe Ia sample~\citep{2022ApJ...938..113S}. A similar approach is found in~\cite{2024MNRAS.533..743W}, but considering the $w_{0}w_{a}$CDM model. The $w_{0}w_{a}$CDM model, known also as the Chevallier-Polarski-Linder (CPL) parametrization~\citep{doi:10.1142/S0218271801000822,PhysRevLett.90.091301}, considers a varying Energy equation-of-state (EoS) where $w$ is parametrized as an evolving function of redshift.

An important question to answer is to what extent the precision on the cosmological parameter $w$ can be expected from the new observations of the GRBs. To tackle this goal, we need to have the smallest possible scatter in the GRB 2D correlations. We estimate the number considering the data that is currently available and with the data expected to be gathered in the next few years by present and future deep-space satellite missions and ground-based facilities. Additionally, we discuss how recent developments in the machine learning techniques applied to GRBs can improve our results.

We follow a similar approach to the one in~\cite{DainottiVia2022MNRAS.514.1828D} in which forecasts were adopted. 
However, this work differs from previous GRB cosmology analyses by forecasting the precision that can be achieved with additioal next-generation GRB plateau samples and by estimating the number of sources required to reach Planck-level ~\citep{planck2018} constraints within the $w$CDM and $w_0w_a$CDM frameworks.
The main goals of this work are threefold. First, we evaluate the precision with which GRBs alone can constrain the Dark EoS parameter $w$, using simulated samples constructed from the observed plateau correlations. Second, we investigate how the combination of GRB data with CMB observations improves cosmological parameter constraints, and we estimate the GRB sample size required to reach the precision currently achieved by experiments such as Planck and DESI. Third, we explore the potential of future GRB samples to test DDE scenarios within the $w_{0}w_{a}$CDM framework, assessing whether GRB observations can provide meaningful constraints on the evolution of the Dark Energy EoS. 
Our results provide a quantitative roadmap for the role of GRBs in future cosmological studies, highlighting how forthcoming missions and improved analysis techniques may transform GRBs into competitive high-redshift probes of Dark Energy.

The paper is structured as follows: in Section \ref{wCDMmodel} we introduce the $w$CDM and $w_{0}w_{a}$CDM models. In Section \ref{datasample} we describe the GRB data set, while in Section \ref{methods} we present the methodology to build simulations, including data from the CMB. In Section \ref{results}, we show the results obtained from the cosmological analysis. In Section \ref{future}, we estimate the future rates of GRB observations. In Section \ref{conclusions}, we provide a summary of the main results and our conclusions.

%%%%%%%%%%%%%%%%%%%%%%%%%%%%
%%%%%%%%%%%%%%%%%%%%%%%%%%%%
%%%%%%%%%%%%%%%%%%%%%%%%%%%%

%%%%%%%%%%%%%%%%%%%%%%%%%%%%
%%%%%%%%%%%%%%%%%%%%%%%%%%%%
%%%%%%%%%%%%%%%%%%%%%%%%%%%%

\section{The $w$CDM, $w_{0}w_{a}$CDM models and the cosmological constraints}\label{wCDMmodel}
In this Section, we introduce the cosmological model used in the present analysis. The $w$CDM model is an extension of the standard $\Lambda$CDM model of cosmology, and to understand its nature, it is crucial to introduce the Equation of state (EoS) in cosmology, a relation between the pressure of the cosmic fluid ($p$) and its energy density ($\rho$): $w=\frac{p}{\rho}$. Based on its value, a given component of the universe is considered to dominate the expansion dynamics. While $\Lambda$CDM assumes that Dark Energy is a cosmological constant (with EoS parameter $w=-1$), the $w$CDM model generalizes this by allowing 
$w$ to be a free parameter. The Hubble parameter $H(z)$ takes the following form:

\begin{equation}
    H(z)=H_0\sqrt{\Omega_{M}(1+z)^3 + \Omega_{r}(1+z)^4 + \Omega_{DE}(1+z)^{3(1+w)}},
    \label{eq:H(z)wCDM}
\end{equation}
where $z$ is the redshift, $\Omega_{M}$ is the total matter (including baryonic matter and Dark Matter) density parameter measured today, $\Omega_{r}$ is the radiation contribution (which can be neglected in the late universe), and $\Omega_{DE}$ is the Dark Energy density parameter responsible for the accelerated expansion of the universe. The $\Omega_{M}$ parameter can be broken down into two main components: the baryonic energy density $\Omega_{b}$ and the Cold Dark Matter energy density $\Omega_{CDM}$, which together satisfy $\Omega_{M}=\Omega_{b}+\Omega_{CDM}$. In the case of the $w_{0}w_{a}$CDM model, the EoS parameter assumes the form $w(z)=w_{0} + w_{a} \frac{z}{1+z}$~\citep{doi:10.1142/S0218271801000822,PhysRevLett.90.091301} and, consequently, the $H(z)$ expression becomes:

\begin{equation}
    H(z)=H_0\sqrt{\Omega_{M}(1+z)^3 + \Omega_{DE}(1+z)^{3(1+w_0+w_a)}e^{-3w_a\frac{z}{1+z}}}.
    \label{eq:H(z)w0waCDM}
\end{equation}

% ACCORDING TO CHATGPT THE FOLLOWING IS A LESS SAFE VERSION OF THE SAME SENTENCE BELOW TO JUSTIFY THE USE OF w0waCDM AT zCMB -> \textcolor{brown}{Although the CPL parametrization is not intended to be extrapolated to CMB redshift, we stress that our analysis focuses on the precision of the inferred cosmological parameters rather than on the physical validity of the dark-energy evolution at high redshift. Therefore, the extrapolation of the CPL form to $z\sim 1100$ should be regarded as a purely phenomenological extension that does not affect the robustness of our uncertainty estimates.}

We note that the CPL parametrization provides a convenient phenomenological description of dark energy at low to intermediate redshift, while its extrapolation to the CMB regime should be interpreted with caution~\citep{PhysRevLett.90.091301}.
However, for the fiducial parameters adopted in this work, $(w_0, w_a)=(-1.09,-0.667)$, the high-redshift limit of the equation of state is $w(z\to\infty)=w_0+w_a<-1$. In this regime, the dark energy density scales as $\rho_{\rm DE}\propto (1+z)^{3(1+w_0+w_a)}$ and therefore decreases rapidly towards the past, becoming negligible compared to matter and radiation densities, that increase as $(1+z)^3$ and $(1+z)^4$. This ensures that the CPL parametrization does not significantly affect the early-Universe physics and can be safely extended to the CMB epoch in our analysis~\citep{PhysRevLett.90.091301}.
The CPL parametrization provides a flexible phenomenological description of a possible redshift evolution of the dark-energy equation of state. In this framework, $w_0$ represents the present-day value of the equation of state, while $w_a$ quantifies its evolution with redshift. Owing to its simplicity and generality, the CPL parametrization is widely used to explore departures from a cosmological constant.
At the same time, constraining the CPL parameters requires careful consideration of the sensitivity of cosmological observables. Most probes do not measure the equation of state directly, but instead constrain integrated quantities of the expansion history, such as the Hubble parameter and luminosity distance. As a result, different combinations of $w_0$ and $w_a$ can produce very similar distance–redshift relations, leading to degeneracies between these parameters~\citep{2014EPJC...74.2729G,lee2025pedagogicnulltestsdynamical}. 
The CMB provides highly precise constraints on cosmological parameters, but it is primarily sensitive to the integrated expansion history up to the last-scattering surface at $z\approx1100$. Within the CPL framework, this leads to constraints on specific combinations of $w_0$ and $w_a$, rather than on each parameter independently. Consequently, complementary probes sampling lower and intermediate redshifts are essential to improve constraints on the evolution of the dark-energy equation of state.
Although extending the redshift coverage of cosmological observations can, in principle, enhance sensitivity to evolving dark-energy models, constraining both $w_0$ and $w_a$ simultaneously remains challenging. This reflects the fact that the CPL parametrization requires precise measurements across a broad and well-sampled redshift range. As we will show in Sec. \ref{results}, this behavior is also observed in our analysis, where the parameter $w_a$ remains less constrained than $w_0$ even for large samples of high-redshift standardizable sources.
These considerations highlight the importance of combining complementary cosmological probes in order to fully exploit the constraining power of the CPL parametrization and to improve constraints on dynamical dark-energy scenarios.
The adoption of a cosmological model allows us to define the luminosity distance as:

\begin{equation}
    D_L(z)=c\,(1+z)\int_{0}^{z}\frac{dz'}{H(z')},
    \label{eq:lumdistance}
\end{equation}
where $c$ is the speed of light in vacuum. The respective theoretical distance modulus is:

\begin{equation}
    \mu_{th}=5\log_{10}[D_L(z)]+25,
    \label{eq:distancemoduli}
\end{equation}
with $D_L$ expressed in megaparsec (Mpc). In the cosmological analysis that involves the use of standard candles, $\mu_{th}$ is compared directly with the observed distance modulus $\mu_{obs}$ of the given probe, in this case, GRBs.

The analysis of CMB fluctuations allows the definition of cosmological observables that provide further constraints on the basic cosmological parameters such as $H_0$, $\Omega_M$, and $w$ in the current work. The first observable to be introduced is the angular size of the sound horizon at the decoupling epoch, $\theta_{MC}$. This is defined as

\begin{equation}
    \theta_{\rm{MC}}=\frac{r_s(z_{LSS})}{D_A(z_{LSS})},
    \label{eq:thetaMC}
\end{equation}
where $r_s(z_{LSS})$ is the comoving sound horizon estimated at the Last Scattering Surface (LSS) $z_{LSS}\sim1100$ and $D_A(z_{LSS})$ is the angular diameter distance to the LSS.
Another relevant observable is the \textit{optical depth} ($\tau$). This expresses the fraction of CMB photons that are scattered by the free electrons during the reionization epoch and can be computed through the integral

\begin{equation}
    \tau=\int_{0}^{z_{RE}}c\, n_e(z)\sigma_T\frac{dt}{dz}dz,
    \label{eq:tau}
\end{equation}
where $z_{RE}$ is the reionization epoch redshift, $n_e(z)$ is the density of free electrons, and $\sigma_T$ is the Thomson cross section.
The amplitude of the primordial scalar perturbations ($A_{\rm s}$) sets the initial amplitude of the scalar density fluctuations that arise during cosmic inflation. It can be connected to the power spectrum of the primordial scalar perturbations $P_s(k)$ through the following:

\begin{equation}
    P_s(k)=A_{\rm s}\biggr(\frac{k}{k_p}\biggr)^{n_s-1},
    \label{eq:As}
\end{equation}
$k$ being the \textit{comoving wave number}, $k_p$ the pivot scale, and $n_s$ the scale dependence of the primordial power spectrum.
All the CMB observables are well constrained by Planck CMB measurements \citep{planck2018}.

\section{The Observed Data Sample}\label{datasample}

We begin our analysis with the full observed sample of X-ray afterglows of purely long GRBs (168 sources with redshift span $0.09\leq z\leq6.29$), excluding SNe Ib/c associated sources, because we have shown in \citealt{Dainotti2017a} that the 2D relation associated with SNe Ib/c follows a steeper relation than the regular long GRBs, thus possibly showing a different emission mechanism or different progenitor system. The full optical sample has 179 sources with $0.06\leq z\leq 8.23$. The X-ray sample consists of sources observed by the Swift satellite from its first observation in January 2005, with redshifts, until December 2023.
The Dainotti relation, both for the X-ray and optical plateaus, is expressed by:

\begin{equation}
    \log_{10} L_{a,i} = a_i \cdot \log_{10} T^{*}_{a,i} + c_i,
    \label{eq:2drelation}
\end{equation}
where $i=X,opt$ denotes the wavelengths, $a_i,c_i$ are the correlation parameters in the $i$ wavelength.
We here assume that the GRB parameters in X-rays and optical are not affected by a significant evolutionary effect with redshift; thus, in the present analysis, no correction is applied to them.
To compare the observed GRB data with the theoretical cosmological predictions, it is crucial to derive from Equation \ref{eq:2drelation} the observed distance modulus ($\mu_{obs}$) for a GRB:

\begin{equation}
    \mu_{obs}=5[a_{1,i} \cdot \log_{10}T^{*}_{a,i} + c_{1,i} + d_{1,i} \cdot \log_{10}F_{a,i}]+25,
    \label{eq:observeddistancemoduli}
\end{equation}
where $F_{a,i}$ is the $k$-corrected end-of-plateau flux in the $i$ wavelength, and the parameters $a_{1,i},c_{1,i},d_{1,i}$ are functions of the parameters $a_{i},c_{i}$, see~\cite{DainottiVia2022MNRAS.514.1828D} for further details. For the likelihood analysis, the intrinsic scatter parameter $\sigma_{GRB}$ is introduced in the total uncertainty of the observed distance moduli, leading to the formula:

\begin{equation}
    \sigma^{2}_{\mu_{obs}}=5^2(a_{1,i} \cdot \sigma_{\log_{10}T^{*}_{a,i}})^2 + 5^2(d_{1,i}\cdot \sigma_{\log_{10}F_{a,i}})^2 + 5^2 \sigma_{GRB,\, i}^{2}.
    \label{eq:totaluncertainty}
\end{equation}

\section{Methodology}\label{methods}

The method is composed of three parts: first, we construct a mock sample; second, we use the simulated data to infer the cosmological parameters through the cosmological fitting procedure; and third, we interpolate the constraints obtained from the fitting procedure. 
The total sample shows a significant scatter in the 2D relation, which can be caused by several factors, including the presence of long GRBs of merger origin ~\citep{Petrosian2024,Lenart2025}.

\subsection{Mock Samples}

\begin{figure}[ht!]
    \centering
    \includegraphics[width=0.49\linewidth]{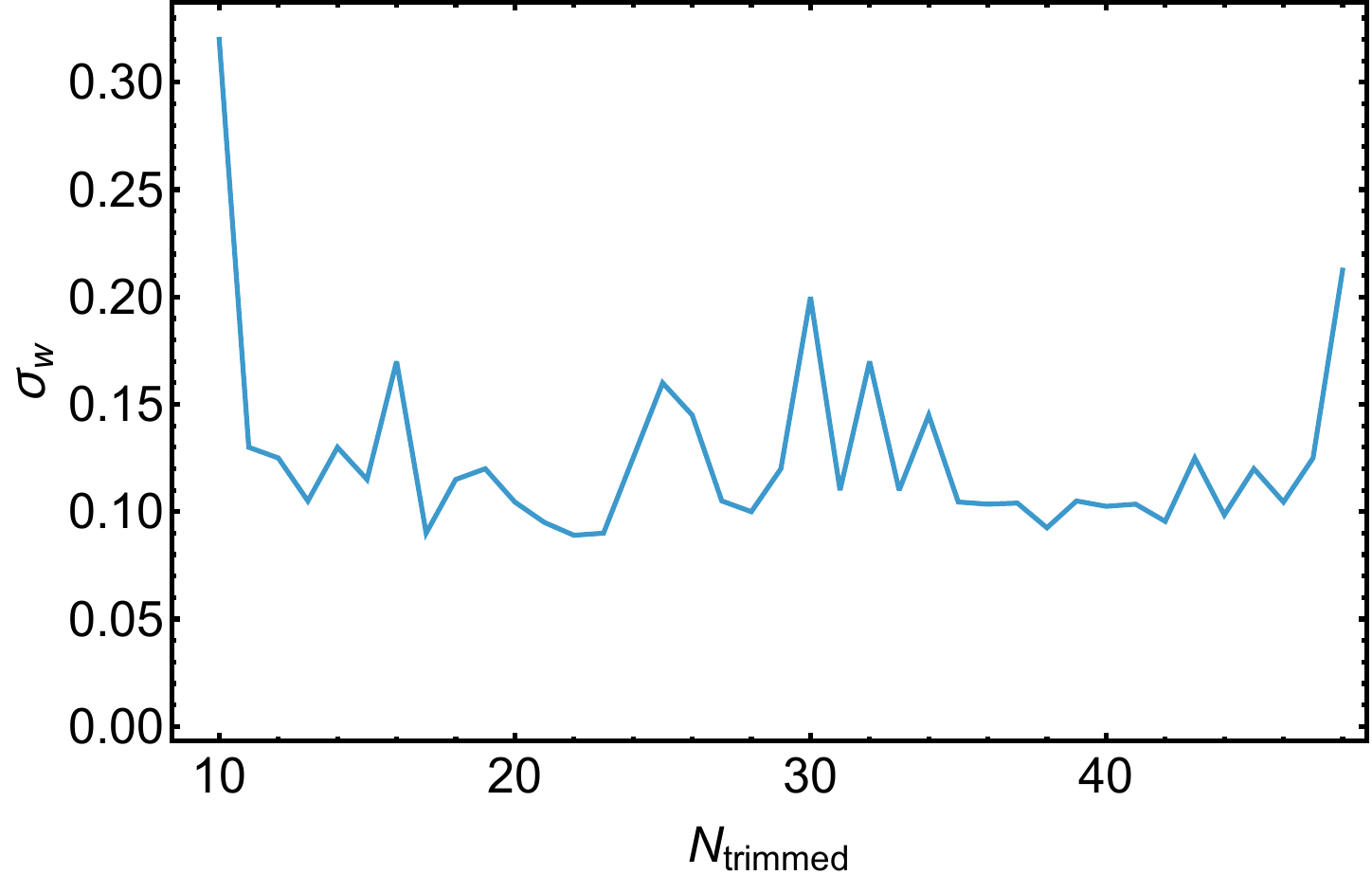}
    \includegraphics[width=0.49\linewidth]{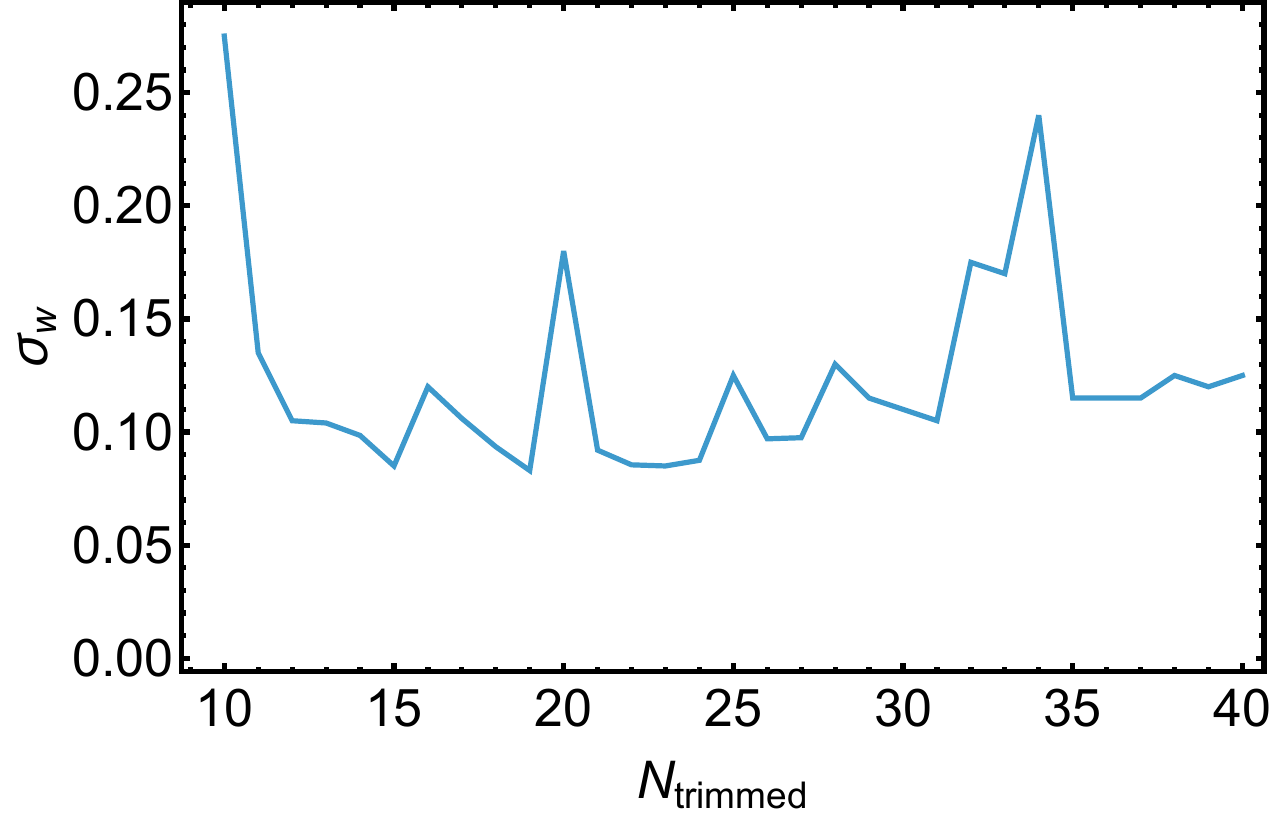}
    \caption{Left: The uncertainty on $w$ derived with mock samples of 2300 GRBs simulated based on the $N_{trimmed}$, the closest GRBs to the best-fit of the X-ray relation. The simulation was based on the varied $a$ $c$ $\sigma_{int}$ parameters taken from the mean fit of the trimmed sample. Right: The same, but for the optical relation.}
    \label{fig:Aposteriori}
\end{figure}

With the sample described above, we have a starting point to obtain higher-precision cosmological constraints, which can be reached via an iterative outlier removal process.
Such a procedure was introduced by~\cite{DainottiVia2022MNRAS.514.1828D} as an a posteriori trimming. Subsequently, samples of sources lying closest to the best fit are taken as a parent sample used to create a mock sample. Next, a cosmological parameter (in this case $w$) is fitted for each of those mock samples. Then, one selects the parent sample with the smallest uncertainty on the studied cosmological parameter.

We computed the luminosity of every GRB under the assumption of a flat $\Lambda$CDM model corresponding to the Planck~\citep{planck2018} best-fit cosmology ($\Omega_M=0.3158$, $H_0=67.32\,{\rm km\,s^{-1}\,Mpc^{-1}}$, $w=-1$). For the extended cosmological models, the GRB mock samples are generated using the corresponding Planck best-fit parameters, with $(w_0,w_a)=(-0.667,-1.09)$ for the $w_0w_a$CDM case. In addition, for part of our analysis, see Figure \ref{fig:w2DMockCMB}, we generate simulated Planck-like CMB likelihoods consistent with the same fiducial cosmologies using the Cobaya-based mock-data framework described in~\cite{Rashkovetskyi2021}. These mock likelihoods allow us to test the joint constraining power of GRBs and CMB data in a controlled setting, free from potential fluctuations or parameter preferences present in the real Planck measurements.

Furthermore, we choose a consecutive subsample of growing size $N_{trim}$ of sources lying closest to the fitting line. Each of these subsamples is the basis to simulate a sample of 2300 sources, and we performed the cosmological fitting of the $w$ parameter. We plot in Figure \ref{fig:Aposteriori} the obtained uncertainty on $w$ as a function of the number of sources in the sample used to create the mock sample for both X-ray and optical data. Therefore, the selected number of sources corresponds to those lying closest to the best-fit correlation line.

We determined the best sample based on this criterion: the largest sample with the smallest uncertainty on $w$, enlarged by 10\% (we use 10\% to account for a small level of randomness in our results, given that multiple values are very close to the minimum). Final samples comprise 42 long GRBs (LGRBs) lying closest to the 2D X-ray relation and 24 GRBs in the optical sample. Then, we used these two sets to simulate high precision GRB samples of increasing size for use in cosmological simulations.

\begin{table}[ht]
\begin{center}
\renewcommand{\arraystretch}{1.5}
\begin{tabular}{c@{\hspace{1 cm}} c}
\hline
\textbf{Parameter}                    & \textbf{Priors interval}\\
\hline\hline
$a_{i}$                        & $(-1.7\,,\,-0.1)$\\
$c_{i}$                        & $(2\,,\,50)$\\
$\sigma_{GRB,i}$                        & $(0\,,\,1)$\\
$\Omega_{\rm b} h^2$         & $[0.005\,,\,0.1]$\\
$\Omega_{\rm c} h^2$       & $[0.001\,,\,0.99]$\\
$100\,\theta_{\rm {MC}}$             & $[0.5\,,\,10]$\\
$\tau$                       & $[0.01\,,\,0.8]$\\
$\log(10^{10}A_{\rm s})$         & $[1.61\,,\,3.91]$\\
$n_s$                        & $[0.8\,,\, 1.2]$\\
$w$                         & $[-4\,,\, 1]$\\
$w_0$                       & $[-4\,,\, 1]$ \\
$w_a$                       & $[-4\,,\, 1]$\\
\hline\hline
\end{tabular}
\end{center}
\caption{List of the parameters used in the MCMC sampling following a \textit{(min, max)} notation. In round brackets, the GRB parameter priors are reported, while in square brackets the priors adopted for the CMB likelihood are shown. %\textcolor{blue}{\textbf{WG: double-check $w_a$!}\textcolor{magenta}{MGD: @Aleksander, can you please comment on the $w_a$?}}\textcolor{red}{[The priors were as above]}
}
\label{tab:priors}
\end{table}

\subsection{Cosmological fitting procedure}

We follow the approach of~\cite{DainottiVia2022MNRAS.514.1828D} and consider the error on the GRB LC parameters as ``halved'', namely, dividing them by $n=2$. Such an outcome is expected with the ML techniques aimed to reconstruct the GRB LCs and improve the precision on the fitting parameters~\citep{2023ApJS..267...42D,2024arXiv241220091M,Kaushal2026}. 
The simulated GRB samples are generated assuming the Planck best-fit $\Lambda$CDM cosmology and are subsequently used in the cosmological parameter inference.
The simulated samples undergo an MCMC analysis for estimating the cosmological parameters.

To this end, we use the publicly available sampler \texttt{COBAYA}~\citep{Torrado:2020dgo}. 
We implemented a dedicated likelihood module for the GRB plateau correlation within the \texttt{Cobaya} framework.
The code explores the posterior distributions of a given parameter space using the MCMC sampler developed for \texttt{CosmoMC}~\citep{Lewis:2002ah}.

Our baseline sampling considers the seven $w$CDM parameters, namely $\omega_{\rm b}\equiv \Omega_{\rm b}h^2$, $\omega_{\rm dm}\equiv\Omega_{\rm CDM} h^2$ ($h=H_0/100km\,s^{-1}\,Mpc^{-1}$ being the dimensionless Hubble parameter), $\theta_{\rm{MC}}$, $\tau$, $\log(10^{10}A_{\rm s})$, $n_s$, and the Dark Energy EoS parameter $w$ or in case of $w_0w_a$CDM model, the CPL EOS parameters $w_0$ and $w_a$~\citep{doi:10.1142/S0218271801000822,PhysRevLett.90.091301}.
In the case of GRBs treated without CMB data, we sample $\Omega_M=\Omega_{\rm b}+\Omega_{\rm dm}$, instead of $\omega_{\rm b}$ and $\omega_{\rm dm}$, since the luminosity distance depends only on the combination of those parameters, as well as $H_0$ and $w$ or $w_0$ and $w_a$. In addition, we consider the GRB luminosity-time relation parameters $a_i,c_i,\sigma_{GRB,i}$.
The prior distributions for the sampled cosmological parameters involved in our analysis are chosen to be uniform, see Table \ref{tab:priors}.

Apart from the sampling cosmological parameters listed above and the nuisance parameters used to model the CMB experiment
systematics and GRBs, we also obtain constraints on some important derived parameters, such as $H_0$ and $\sigma_8$. The convergence of the chains obtained with this procedure is tested using the Gelman-Rubin criterion~\citep{Gelman:1992zz}. 

In all our runs we combine the GRBs described in Section \ref{datasample} with Planck 2018 temperature and polarization (TT TE EE) likelihood, which also includes low multipole data ($\ell < 30$, \citealt{Planck:2019nip,Planck:2018vyg,Planck:2018nkj}) and the Planck 2018 lensing likelihood~\citep{Planck:2018lbu}, constructed from measurements of the power spectrum of the lensing potential. We refer to this dataset as \textit{P18}. 
The Planck likelihood is combined with the simulated GRB samples generated around the Planck best-fit cosmology to forecast the constraining power of future GRB observations. Moreover, for some forecasts we also simulate CMB data generated around the same fiducial cosmologies.

The total likelihood adopted in the current work is given by the combination of the GRB log-likelihood, $\log \mathcal{L}_{GRB}$, and the Planck likelihood, $\log \mathcal{L}_{CMB}$. The first one is defined as follows:

\begin{equation}
\begin{split}
\log \mathcal{L}_{GRB}= & -\frac{1}{2}\biggl[\sum_{i=1}^N\log(\sigma^{2}_{GRB,i}+a^{2}\cdot\sigma^{2}_{\log_{10}T^{*}_{i,a}}+\sigma^{2}_{\log_{10}L_{i,a}})\biggr] \\
 &  -\frac{1}{2}\biggl[\sum_{i=1}^N \frac{(\log_{10}L^{th}_{i,a}-\log_{10}L_{i,a})^2}{(\sigma^{2}_{GRB,i}+a^{2}\cdot\sigma^{2}_{\log_{10}T^{*}_{i,a}}+\sigma^{2}_{\log_{10}L_{i,a}})}\biggr]
\end{split}
\label{eq:GRBlike}
\end{equation}

where $L^{th}_{i,a}$ is the vector of theoretical end-of-plateau luminosities computed through the relation in Equation \ref{eq:2drelation}. The CMB likelihood, instead, can be schematically written as

\begin{equation}
    \log \mathcal{L}_{CMB} \propto -\frac{1}{2}(D-M(\eta))^{T}\mathcal{C}^{-1}(D-M(\eta)),
    \label{eq:CMBlike}
\end{equation}
where $D$ is the vector of the measured CMB power spectra, $M(\eta)$ is the model prediction based on the $\eta$ vector of parameters, and $\mathcal{C}$ is the covariance matrix that includes both statistical and systematic contributions. In practice, however, the CMB likelihood is evaluated using the full Planck 2018 likelihood implemented in \texttt{Cobaya}, while in the forecast case simulated CMB spectra generated around the chosen fiducial cosmology are used instead.

\subsection{Interpolation of the results}

\begin{figure}
    \centering
    \includegraphics[width=0.49\linewidth]{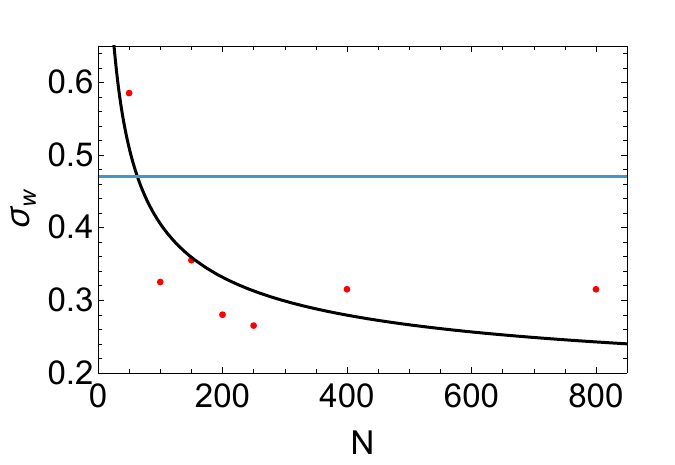}
    \includegraphics[width=0.49\linewidth]{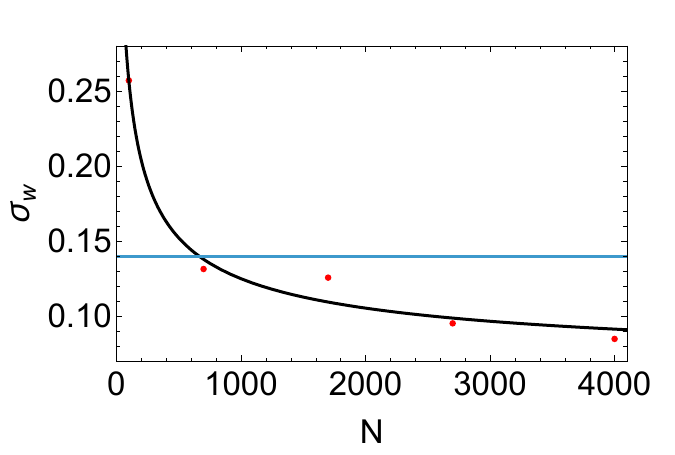}
    \caption{Uncertainty on the $w$ parameter as a function of the GRB sample size used for computation. The blue continuous line is the uncertainty reached with Planck alone (left panel) and DESI alone (right panel).}
    \label{fig:werrorvsN}
\end{figure}

%Our simulations require computational time. Therefore, to estimate the number of sources required to reach a given precision, we carried out part of the simulations on Cray XD2000 at the Center for Computational Astrophysics, National Astronomical Observatory of Japan, we interpolate between the data points. 
Our simulations are computationally expensive. Therefore, to estimate the number of sources required to reach a given precision, we interpolate between the simulated data points. Part of the simulations was carried out on the Cray XD2000 at the Center for Computational Astrophysics, National Astronomical Observatory of Japan, and part on the Stanage cluster at the University of Sheffield.
We choose a specific interpolation function relating the uncertainty in a given parameter to the number of GRBs $N$. When combining independent measurements (calculating the mean weighted by the inverse square of errors) of a single parameter, the resulting variance satisfies the inverse-variance relation. In our case, combining Planck and GRB data, the resulting uncertainty can be approximated by the formula:

\begin{equation}
\sigma_{w_{\#},\,GRB+Planck} =
\sqrt{
\frac{1}{
\frac{1}{\sigma_{w_{\#},\, Planck}^2}
+
\frac{N}{n}\frac{1}{\sigma_{w_{\#},\, GRB}^2}}
}
,
\end{equation}
where $\sigma_{w_{\#},\, Planck}$ is the uncertainty on $w_{\#}$ ($w$ or $w_0$ or $w_a$) from the Planck measurements only and $\sigma_{w_{\#},\, GRB}$ is the uncertainty on $w_{\#}$ computed with a sample of $n$ sources, which create a sample with parameter distributions representative of the parent population of $N$ sources. The factor $N/n$ represents the number of statistically independent subsamples contained in the total GRB population. However, this is an overly idealized model. The true uncertainty in our case depends on the posterior probability in the multidimensional parameter space. Thus, we generalize and simplify the above formula to use it as an interpolation function in our empirical simulations:

\begin{equation}
\sigma_{w_{\#},\,GRB+Planck}(N) = \alpha \sqrt{\frac{1}{\beta+N}}+\gamma,
\end{equation}
where the $\alpha$, $\beta$ and $\gamma$ are the free parameters of the least squares fitting. We found that this function fits most of the results of our simulations. Therefore, we use it when possible, and in the other cases we use polynomials. For the relation between $w$ or $w_0$ or $w_a$ and the sample size $N$, we interpolate using polynomials. The example of the uncertainty interpolation is shown in Figure \ref{fig:werrorvsN}.

\section{Results}\label{results}
The results are divided into three parts: all of them deal with simulated GRB data. The first part presents the GRB analysis and evaluates the number of GRBs required to reach a precision comparable to that of Planck. The second part considers the combination of GRBs with Planck data, either simulated or observed. Both sections aim to check to what extent we can constrain the $w$ parameter. The last section combines simulated Planck data with GRBs to test the constraining power of these combined data sets for the $w_0w_a$CDM model.

\subsection{The precision of Planck on the wCDM model with standalone GRBs}

\begin{table*}
\setlength{\tabcolsep}{0pt}

\scriptsize

\centering
\renewcommand{\arraystretch}{2}
\caption{Summary of the forecast scenarios considered in this work. The table reports the cosmological model, the parameters varied or fixed in the analysis, the fiducial values used to generate the mock data, and the goal of each test. The last three columns list the number of GRBs required for varying all the parameters of the correlation (vary) / fixing the normalisation (fix $c$) / fixing all parameters of the correlation (calib) strategies, and the corresponding year assuming the current GRB detection rate, together with the expected timeline when using transfer learning techniques.}
\label{tab:results_rotated}

\begin{tabular}{c|c|c|c|c|c|c|c|c|c}
\toprule
Data & Corr. & \shortstack{Parameters\\ varied} & \shortstack{Parameters\\ fixed} & Mock values & Fig. & Goal & \shortstack{\# of GRBs\\ needed} & Year & \shortstack{Year with\\ transfer learning} \\
\midrule

\multirow{10}{*}{Without Planck}
& \multirow{4}{*}{X-ray} & $w$ & - & $w=-1$
& \ref{fig:w2D}
& $w_{\rm Planck}=-1.58^{+0.52}_{-0.41}$
& \shortstack{$864$ (vary)\\$175$ (fix c)\\$93$ (calib)}
& \shortstack{2037\\2026\\now} & \shortstack{2031\\now\\now} \\\cline{3-10}

&  & $w$ & \shortstack{$\Omega_M=0.305$\\ $H_0=68.15\rm\, \frac{km}{s\, Mpc}$} & $w=-1$
& \ref{fig:w2D}
& $w_{\rm Planck}=-1.58^{+0.52}_{-0.41}$
& \shortstack{$158$ (vary)\\$45$ (fix c)\\$12$ (calib)}
& \shortstack{now\\now\\now} & \shortstack{now\\now\\now}  \\\cline{3-10}

& & $w_0,w_a$ & - & \shortstack{$w_0=-0.667$\\ $w_a=-1.09$}
& --
& \shortstack{$w_a+3\sigma_{w_a}<0$ \&\\ $w_0-3\sigma_{w_0}>-1$}
& \multirow{3}{*}{$>4000$}
& \multirow{3}{*}{$>2057$} & \multirow{3}{*}{$>2057$}  \\\cline{3-7}

& & $w_0$ & $w_a=-1.09$ & \shortstack{$w_0=-0.667$\\ $w_a=-1.09$}
& --
& $w_0-3\sigma_{w_0}>-1$
& 
&  &   \\\cline{3-7}

& & $w_a$ & $w_0=-0.667$ & \shortstack{$w_0=-0.667$\\ $w_a=-1.09$}
& --
& $w_a+3\sigma_{w_a}<0$
& 
&  &  \\\cline{2-10}

& \multirow{4}{*}{Opt} 
& $w$ & - & $w=-1$
& \ref{fig:w2D}
& $w_{\rm Planck}=-1.58^{+0.52}_{-0.41}$
& \shortstack{$723$ (vary)\\$80$ (fix c)\\$66$ (calib)}
& \shortstack{2036\\2026\\2026} & \shortstack{2030\\now\\now}  \\\cline{3-10}

&  & $w$ & \shortstack{$\Omega_M=0.305$\\ $H_0=68.15\rm\, \frac{km}{s\, Mpc}$} & $w=-1$
& \ref{fig:w2D}
& $w_{\rm Planck}=-1.58^{+0.52}_{-0.41}$
& \shortstack{$114$ (vary)\\$29$ (fix c)\\$<10$ (calib)}
& \shortstack{2026\\now\\now} & \shortstack{now\\now\\now}  \\\cline{3-10}

& & $w_0,w_a$ & - & \shortstack{$w_0=-0.667$\\ $w_a=-1.09$}
& --
& \shortstack{$w_a+3\sigma_{w_a}<0$ \&\\ $w_0-3\sigma_{w_0}>-1$}
& \multirow{3}{*}{$>4000$}
& \multirow{3}{*}{$>2057$} & \multirow{3}{*}{$>2057$}  \\\cline{3-7}

& & $w_0$ & $w_a=-1.09$ & \shortstack{$w_0=-0.667$\\ $w_a=-1.09$}
& --
& $w_0-3\sigma_{w_0}>-1$
& 
&  &  \\\cline{3-7}

& & $w_a$ & $w_0=-0.667$ & \shortstack{$w_0=-0.667$\\ $w_a=-1.09$}
& --
& $w_a+3\sigma_{w_a}<0$
& 
&  &   \\\cline{1-10}

\midrule

\multirow{2}{*}{With real Planck}
& X-ray & $w$ & - & $w=-1$
& \ref{fig:w2DCMB}
& $w_{\rm DESI}=-0.99^{+0.15}_{-0.13}$
& \shortstack{$921$ (vary)\\$1376$ (fix c)\\$979$ (calib)}
& \shortstack{$2037$\\$2041$\\$2038$} & \shortstack{$2032$\\$2035$\\$2032$}  \\\cline{2-10}

& Opt & $w$ & - & $w=-1$
& \ref{fig:w2DCMB}
& $w_{\rm DESI}=-0.99^{+0.15}_{-0.13}$
& \shortstack{$1410$ (vary)\\$1344$ (fix c)\\$1062$ (calib)}
& \shortstack{$2045$\\$2044$\\$2041$} & \shortstack{$2035$\\$2035$\\$2033$}  \\\cline{1-10}

\midrule

\multirow{8}{*}{With mock Planck}
& \multirow{4}{*}{X-ray}
& $w$ & - & $w=-1$
& \ref{fig:w2DMockCMB}
& $w_{\rm DESI}=-0.99^{+0.15}_{-0.13}$
& \shortstack{$1631$ (vary)\\$1473$ (fix c)\\$666$ (calib)}
& \shortstack{$2043$\\$2042$\\$2034$} & \shortstack{$2037$\\$2036$\\$2030$}  \\\cline{3-10}

& & $w_0,w_a$ & - & \shortstack{$w_0=-0.667$\\ $w_a=-1.09$}
& \ref{fig:w0wa}
& \shortstack{$w_a+3\sigma_{w_a}<0$ \&\\ $w_0-3\sigma_{w_0}>-1$}
& $>4000$
& $>2057$ & $>2057$ \\\cline{3-10}

& & $w_0$ & $w_a=-1.09$ & \shortstack{$w_0=-0.667$\\ $w_a=-1.09$}
& --
& $w_0-3\sigma_{w_0}>-1$
& $>4000$
& $>2057$ & $>2057$  \\\cline{3-10}

& & $w_a$ & $w_0=-0.667$ & \shortstack{$w_0=-0.667$\\ $w_a=-1.09$}
& --
& $w_a+3\sigma_{w_a}<0$
& \shortstack{$2340$ (vary)\\$2399$ (fix c)\\$1793$ (calib)}
& \shortstack{$2052$\\$2053$\\$2045$} & \shortstack{$2039$\\$2041$\\$2038$}  \\\cline{2-10}

& \multirow{4}{*}{Opt}
& $w$ & - & $w=-1$
& \ref{fig:w2DMockCMB}
& $w_{\rm DESI}=-0.99^{+0.15}_{-0.13}$
& \shortstack{$1358$ (vary)\\$1320$ (fix c)\\$1052$ (calib)}
& \shortstack{$2045$\\$2044$\\$2041$} & \shortstack{$2035$\\$2035$\\$2033$}  \\\cline{3-10}

& & $w_0,w_a$ & - & \shortstack{$w_0=-0.667$\\ $w_a=-1.09$}
& \ref{fig:w0wa}
& \shortstack{$w_a+3\sigma_{w_a}<0$ \&\\ $w_0-3\sigma_{w_0}>-1$}
& $>4000$
& $>2057$ & $>2057$ \\\cline{3-10}

& & $w_0$ & $w_a=-1.09$ & \shortstack{$w_0=-0.667$\\ $w_a=-1.09$}
& --
& $w_0-3\sigma_{w_0}>-1$
& $>4000$
& $>2057$ & $>2057$  \\\cline{3-10}

& & $w_a$ & $w_0=-0.667$ & \shortstack{$w_0=-0.667$\\ $w_a=-1.09$}
& --
& $w_a+3\sigma_{w_a}<0$ 
& \shortstack{$1594$ (vary)\\$1513$ (fix c)\\$1329$ (calib)}
& \shortstack{$2053$\\$2048$\\$2044$} & \shortstack{$2037$\\$2036$\\$2035$}  \\

\bottomrule
\end{tabular}

\label{tab:results}
\end{table*}

It is important to stress that the focus of this work is to estimate how many GRBs are required to reach a given precision on the $w$ parameter, rather than to compare the inferred value of $w$ with those currently known values in the literature.
We summarize all results in Table~\ref{tab:results}.

\begin{figure*}[ht!]
\centering
\subfloat[All parameters of the X-ray correlation are varied; all cosmological parameters are free to vary.]{\includegraphics[width=2in]{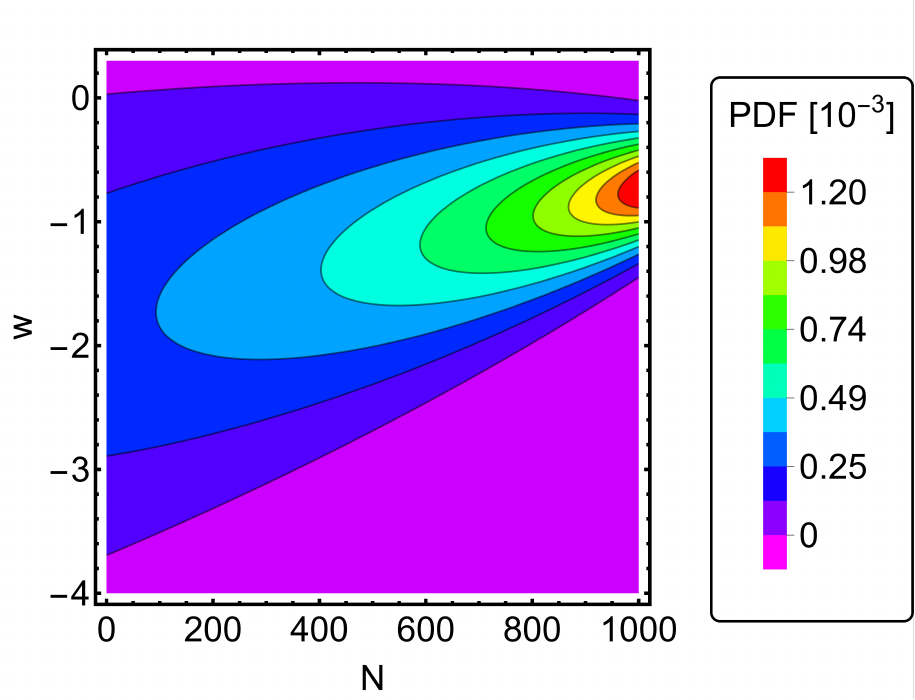}}
\textcolor{blue}{\hspace{0.35cm}}
\subfloat[The normalisation of the X-ray correlation is fixed; all cosmological parameters are free to vary.]{\includegraphics[width=2in]{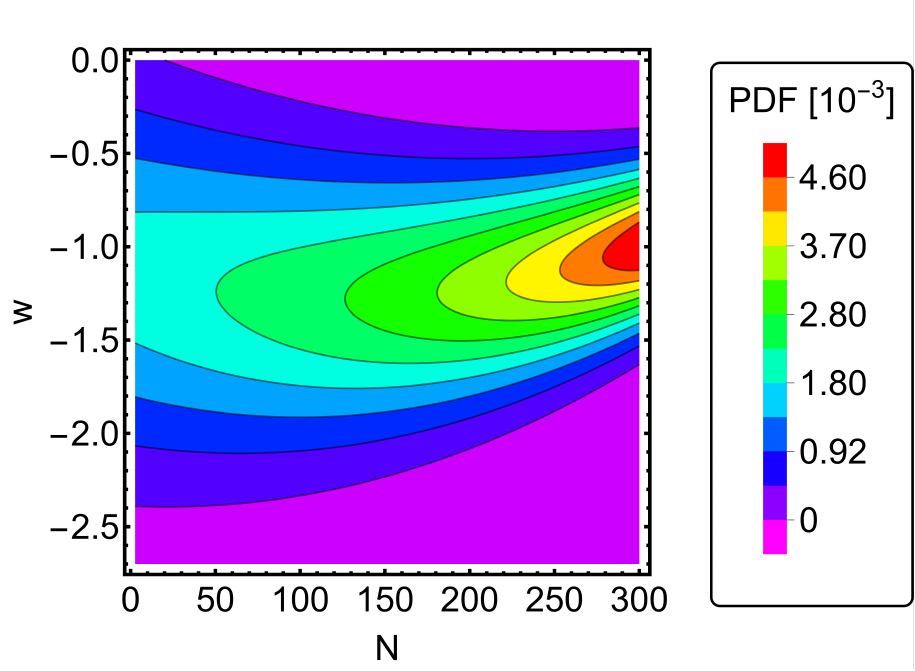}}
\textcolor{blue}{\hspace{0.35cm}}
\subfloat[All parameters of the X-ray correlation are fixed; all cosmological parameters are free to vary.]{\includegraphics[width=2in]{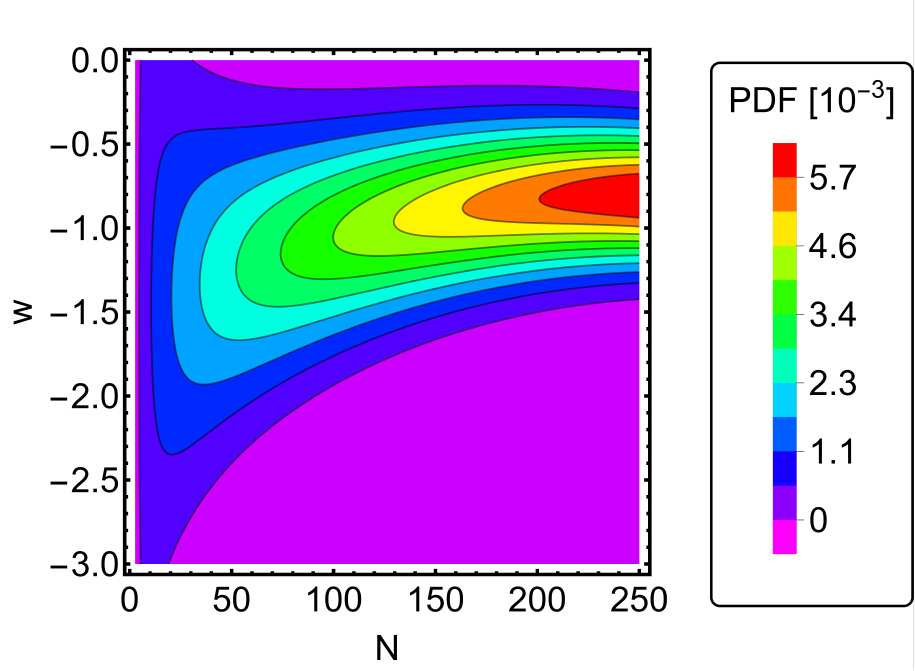}}\\

\subfloat[All parameters of the optical correlation are varied; all cosmological parameters are free to vary.]{\includegraphics[width=2in]{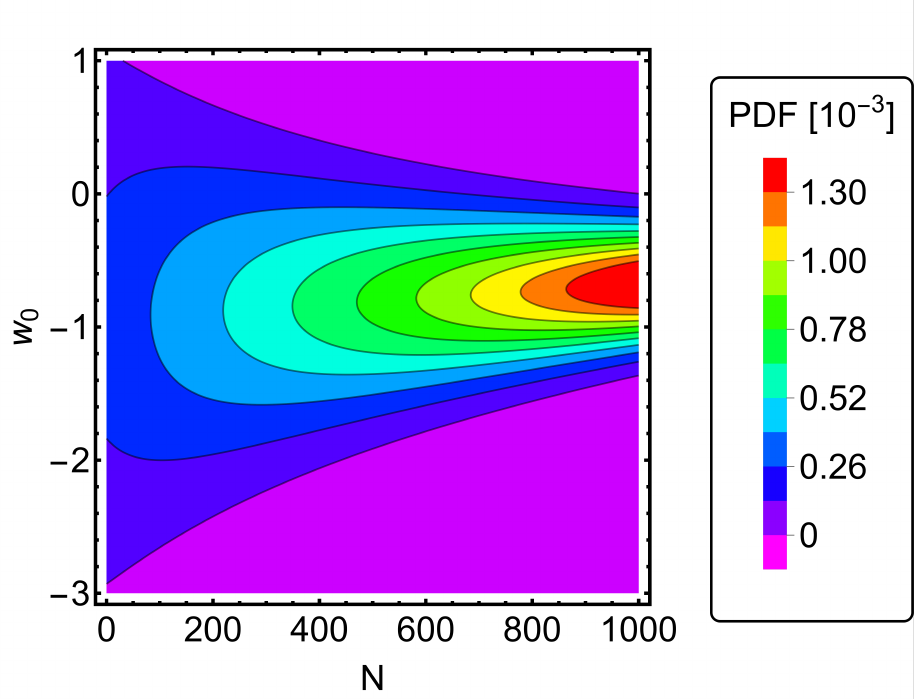}}
\textcolor{blue}{\hspace{0.35cm}}
\subfloat[The normalisation of the optical correlation is fixed; all cosmological parameters are free to vary.]{\includegraphics[width=2in]{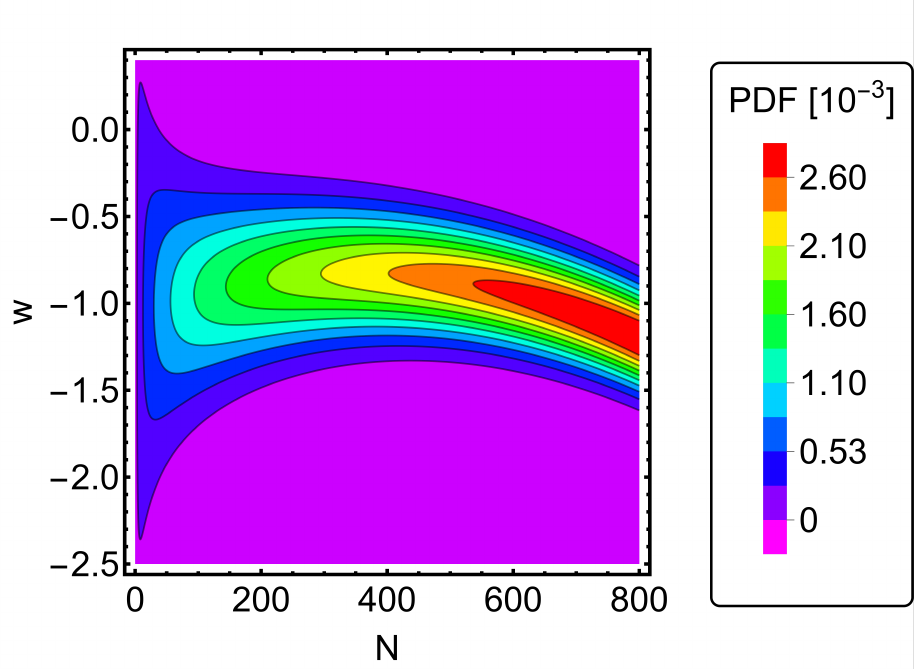}}
\textcolor{blue}{\hspace{0.35cm}}
\subfloat[All parameters of the optical correlation are fixed; all cosmological parameters are free to vary.]{\includegraphics[width=2in]{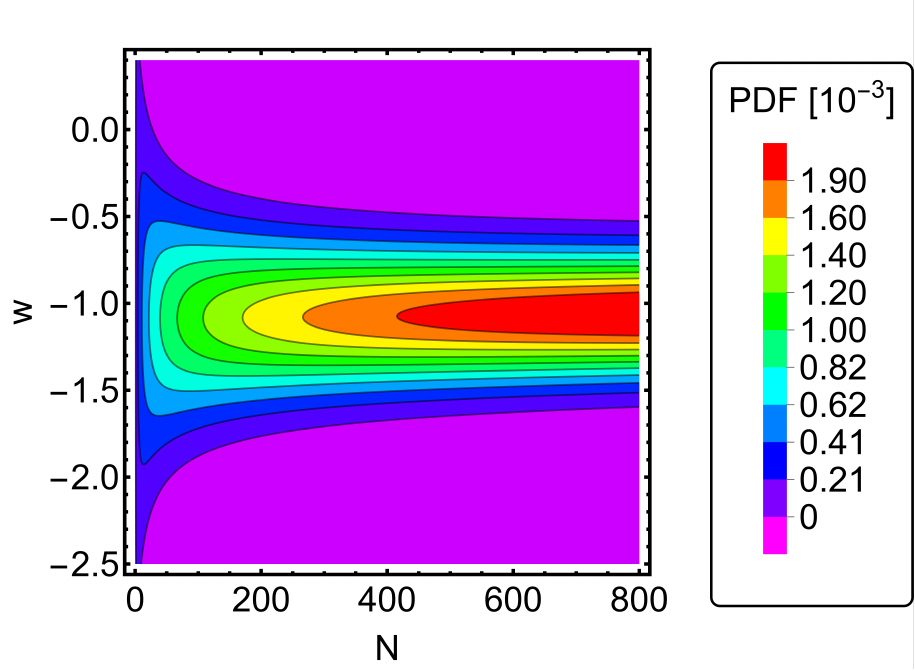}}\\

\subfloat[All parameters of the X-ray correlation are varied; $\Omega_M$ and $H_0$ are fixed.]{\includegraphics[width=2in]{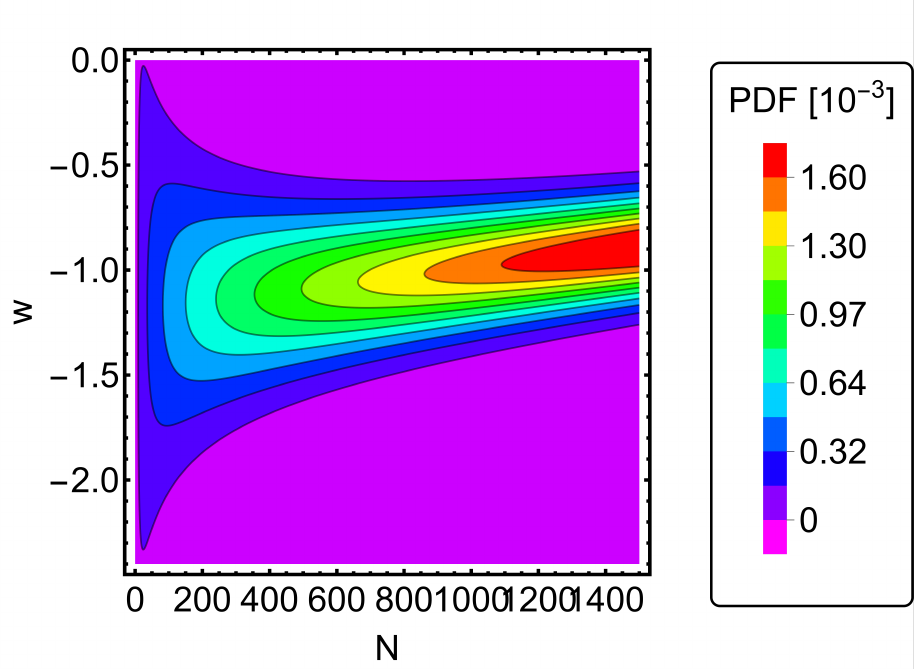}}
\textcolor{blue}{\hspace{0.35cm}}
\subfloat[The normalisation of the X-ray correlation is fixed; $\Omega_M$ and $H_0$ are fixed.]{\includegraphics[width=2in]{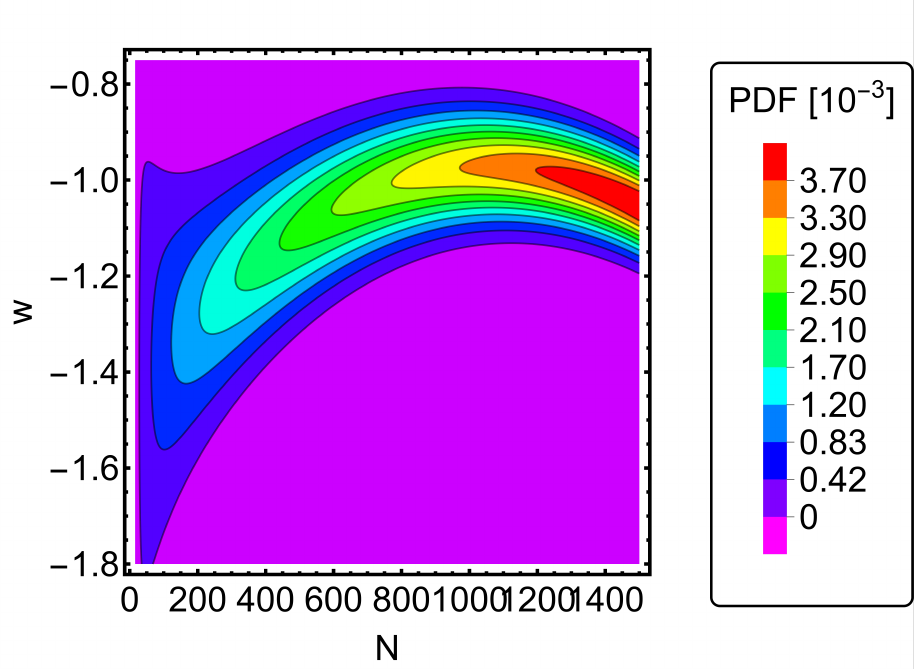}}
\textcolor{blue}{\hspace{0.35cm}}
\subfloat[All parameters of the X-ray correlation are fixed; $\Omega_M$ and $H_0$ are fixed.]{\includegraphics[width=2in]{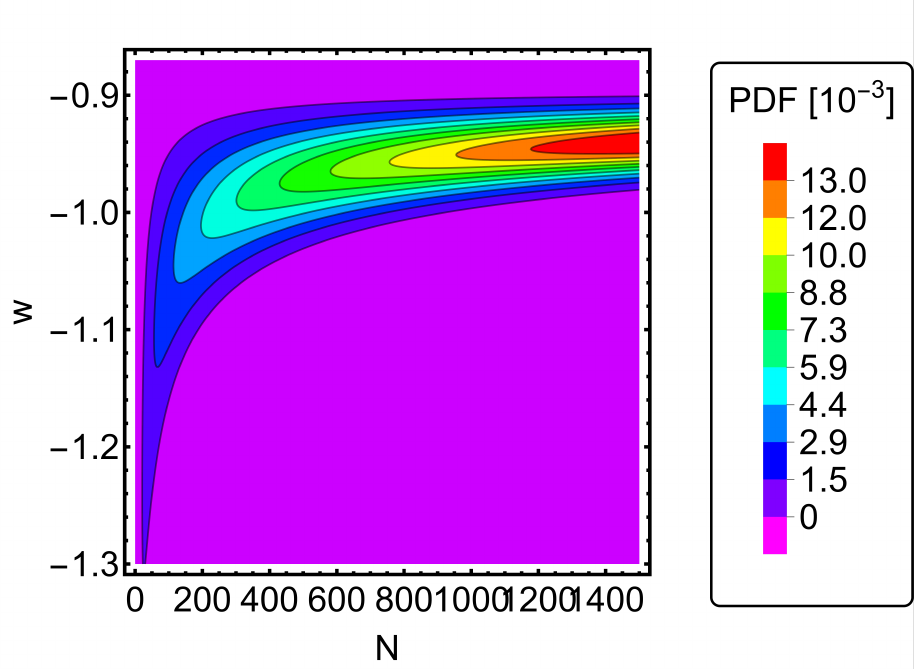}}\\

\subfloat[All parameters of the optical correlation are varied; $\Omega_M$ and $H_0$ are fixed.]{\includegraphics[width=2in]{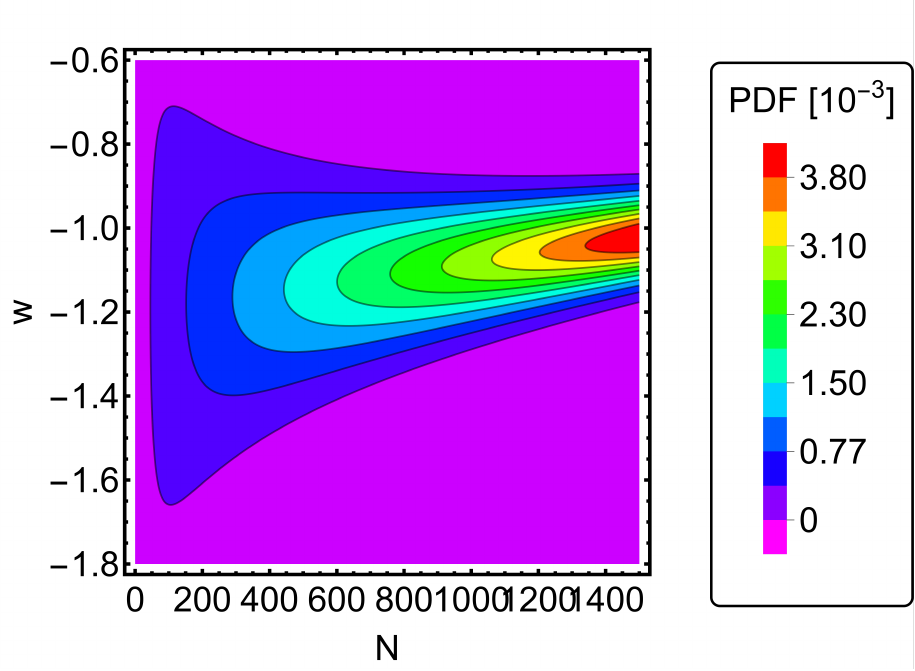}}
\textcolor{blue}{\hspace{0.35cm}}
\subfloat[The normalisation of the optical correlation is fixed; $\Omega_M$ and $H_0$ are fixed.]{\includegraphics[width=2in]{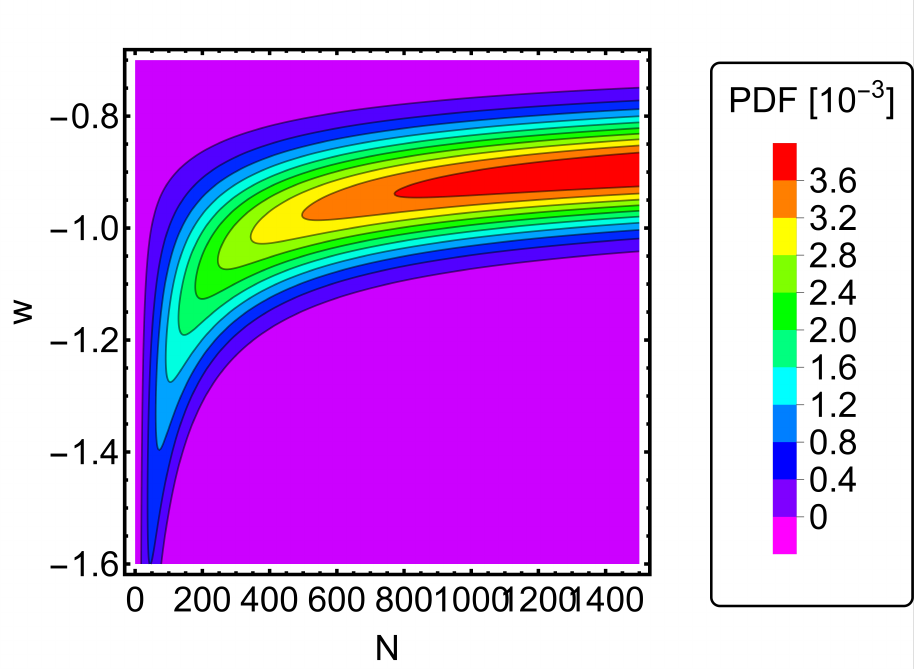}}
\textcolor{blue}{\hspace{0.35cm}}
\subfloat[All parameters of the optical correlation are fixed; $\Omega_M$ and $H_0$ are fixed.]{\includegraphics[width=2in]{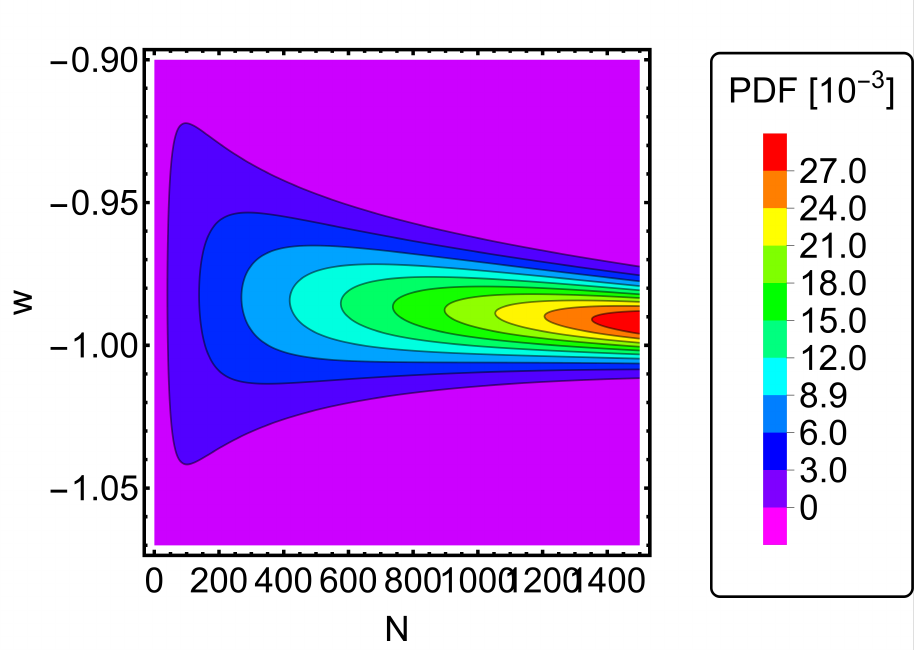}}

\caption{Forecast constraints on $w$ obtained from mock GRB samples. The first two rows correspond to the X-ray and optical correlations analysed with all cosmological parameters free to vary. The last two rows show the corresponding analyses with $\Omega_M$ and $H_0$ fixed. In each case we consider three calibration strategies for the GRB Dainotti 2D correlation: varying all correlation parameters (left column), fixing the normalisation (middle column), and fixing the full correlation (right column).}
\label{fig:w2D}
\end{figure*}

Starting from a sample of $N$ simulated GRBs with a plateau, $N$ is gradually increased, and the precision on $w$ is estimated for each sample of size $N$. $N$ is increased until we reach the precision on $w$ reported in~\cite{planck2018}: $w=-1.58^{+0.52}_{-0.41}$. As a threshold, we here consider the symmetrized uncertainty, namely, $\sigma_w=0.47$.

In the case of the X-ray (optical) part of the analysis, our results indicate that a high-quality sample of 864 (723) GRBs is sufficient to reach the precision of Planck on the $w$ parameter. When we consider the GRB relation with fixed normalisation, this number drops to 175 (80), and in the case where all parameters of the GRB correlation are fixed, we need just 93 (66) sources. 

As shown in Figure~\ref{fig:w2D}, in all those cases the accuracy of $w$ increases as the number of sources increases, although in the case where all GRB parameters are varied, the value is underestimated for a small number of GRBs.

In the case of fixed normalisation and all parameters fixed, the best-fit value is close to the assumed fiducial value $w=-1$ even for a small number of sources.

\subsection{The precision of DESI on the wCDM model with GRBs and Planck together}

\begin{figure*}[ht!]
\centering
\subfloat[All parameters of the X-ray correlation are varied.]{\includegraphics[width=2in]{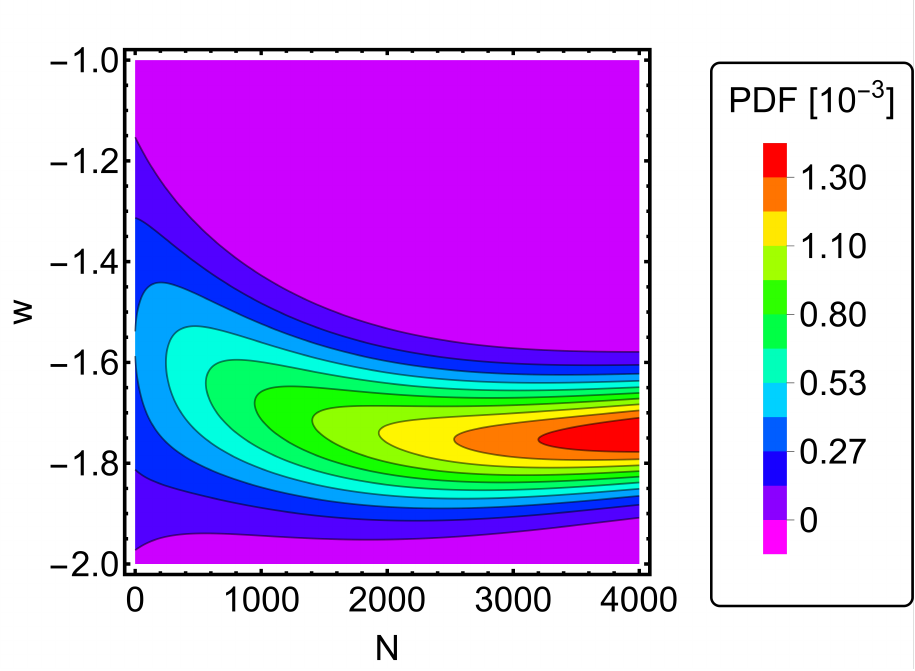}}
\textcolor{blue}{\hspace{0.35cm}}
\subfloat[The normalisation of the X-ray correlation is fixed.]{\includegraphics[width=2in]{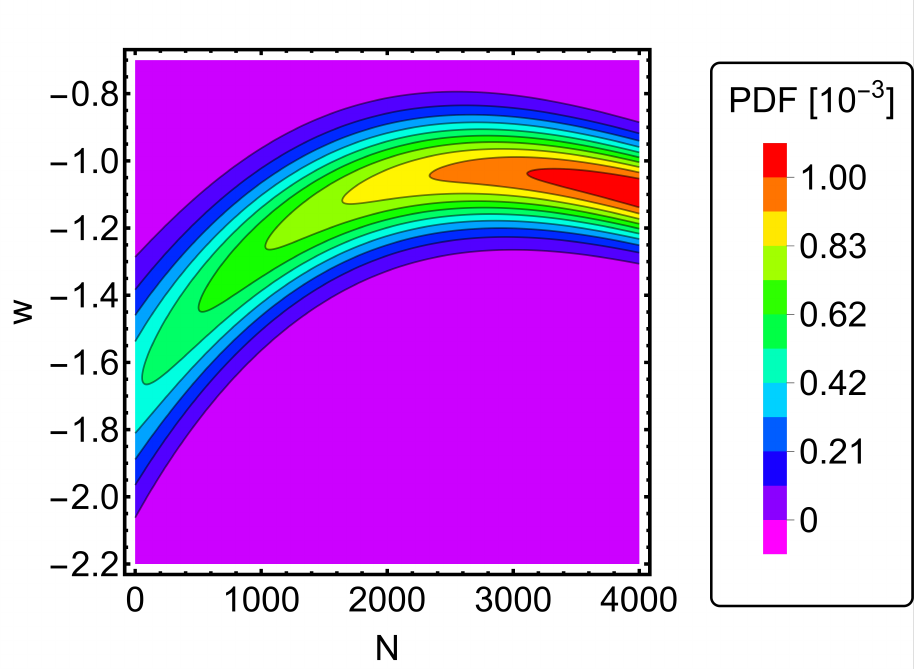}}
\textcolor{blue}{\hspace{0.35cm}}
\subfloat[All parameters of the X-ray correlation are fixed.]{\includegraphics[width=2in]{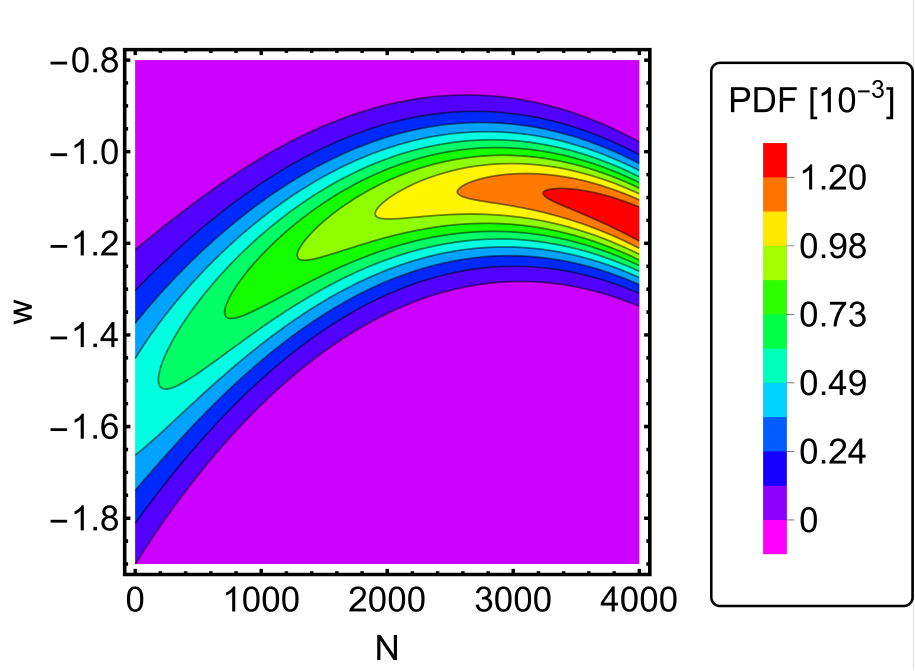}}\\

\subfloat[All parameters of the optical correlation are varied.]{\includegraphics[width=2in]{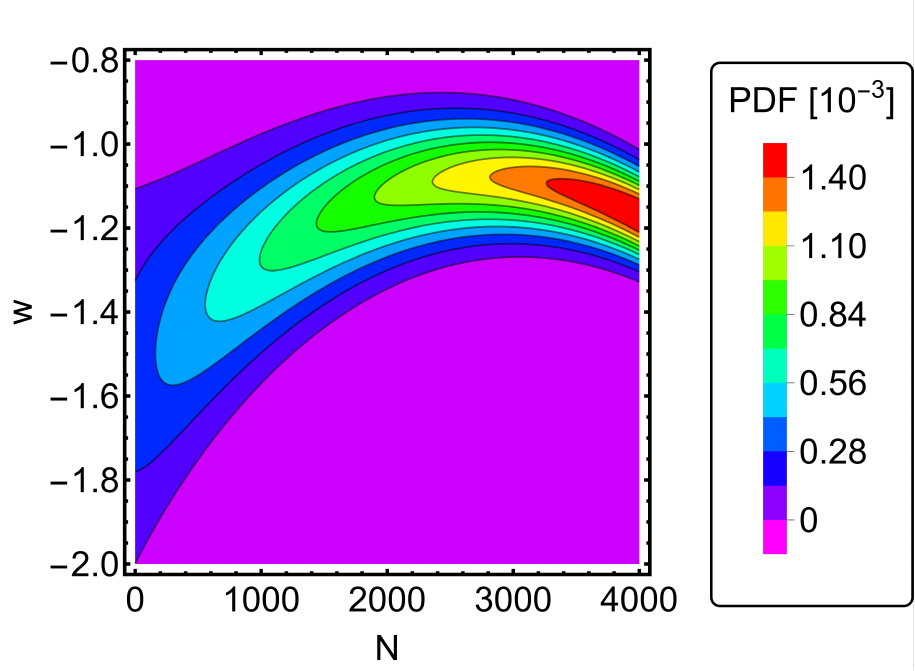}}
\textcolor{blue}{\hspace{0.35cm}}
\subfloat[The normalisation of the optical correlation is fixed.]{\includegraphics[width=2in]{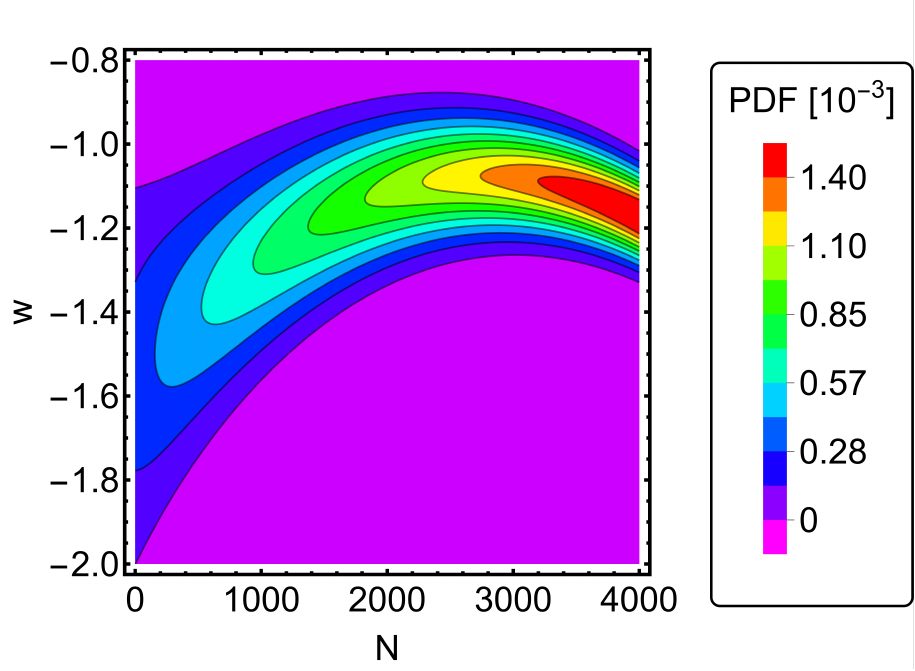}}
\textcolor{blue}{\hspace{0.35cm}}
\subfloat[All parameters of the optical correlation are fixed.]{\includegraphics[width=2in]{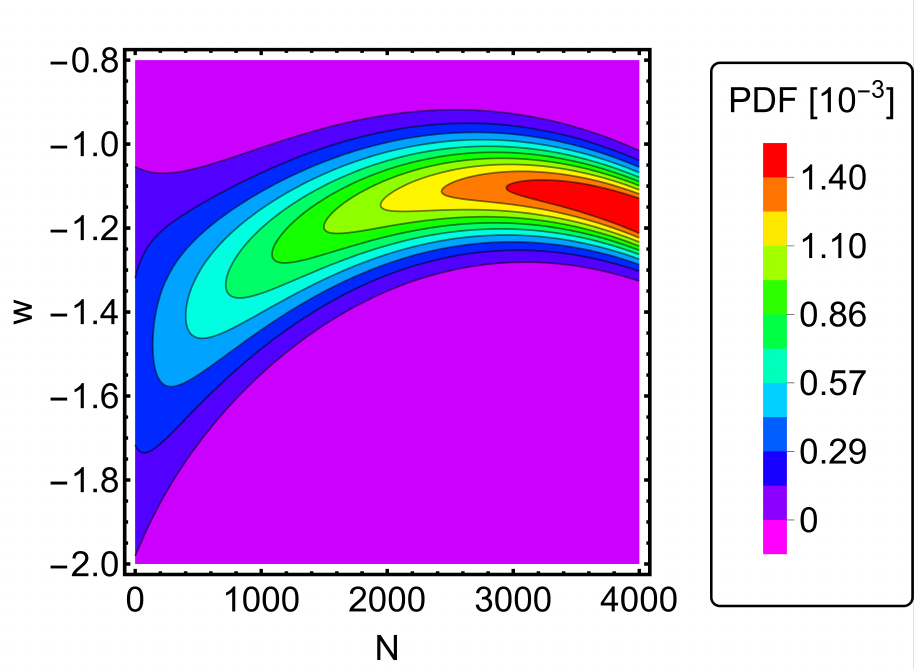}}

\caption{2D X-ray (top row) and optical (bottom row) mock GRBs analysed with the real Planck data.}
\label{fig:w2DCMB}
\end{figure*}

When the X-ray (optical) sample is combined with the real Planck 2018 likelihood, one needs 921 (1410) high-quality GRBs to reach the uncertainty of $0.14$ (symmetrized error reported by \citealt{DESI2025}). When the normalisation of the correlation is fixed, the number changes to 1376 (1344). Thus, for the X-ray sample, the number has increased instead of decreasing. Further, if all the parameters of the GRB correlation are fixed, one needs 979 (1062). 

%It is worth noting that, unexpectedly, varying all parameters of the X-ray correlation appears to be the most efficient strategy. However, as visible in Figure~\ref{fig:w2DCMB}, $w$ is shifted towards smaller values. We suspect that this is due to the well-known geometrical degeneracies present in CMB data within the $w$CDM model. When the number of free parameters is increased, the chains are more strongly driven by the Planck likelihood. This behaviour is mainly related to degeneracies between $w$, $\Omega_M$, and $H_0$ in CMB-only constraints. 

%To avoid these degeneracies in the real Planck data affecting our forecast, we also use simulated CMB data. Within the $w$CDM model, we assume $w=-1$ and compute the mock data with the open-access code~\citep{Rashkovetskyi2021}.

It is worth noting that, somewhat unexpectedly, varying all parameters of the X-ray correlation appears to be the most efficient strategy. However, as shown in Figure~\ref{fig:w2DCMB}, this case also shifts the inferred value of $w$ towards smaller values. This behaviour is likely related to the well-known geometrical degeneracies affecting CMB constraints in the $w$CDM model, in particular those involving $w$, $\Omega_M$, and $H_0$. When additional free parameters are introduced in the GRB sector, the fit becomes more sensitive to these underlying CMB degeneracies, which can in turn affect the apparent constraining power of the combined analysis.

To assess the GRB+CMB constraining power in a more controlled forecasting setup, we therefore also consider simulated CMB data. Within the $w$CDM model, we assume a fiducial $\Lambda$CDM cosmology with $w=-1$ and generate mock CMB data using the public framework described in~\cite{Rashkovetskyi2021}. This allows us to isolate the forecasting performance of the combined analysis without the impact of specific fluctuations or parameter preferences present in the real Planck data.

\begin{figure*}[ht!]
\centering
\subfloat[All parameters of the X-ray correlation are varied.]{\includegraphics[width=2in]{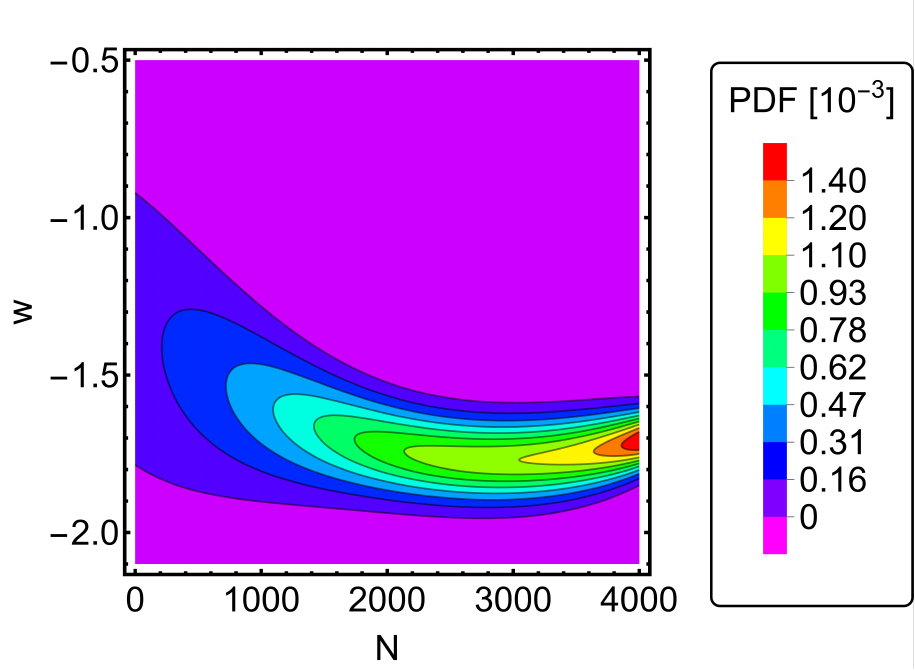}}
\textcolor{blue}{\hspace{0.35cm}}
\subfloat[The normalisation of the X-ray correlation is fixed.]{\includegraphics[width=2in]{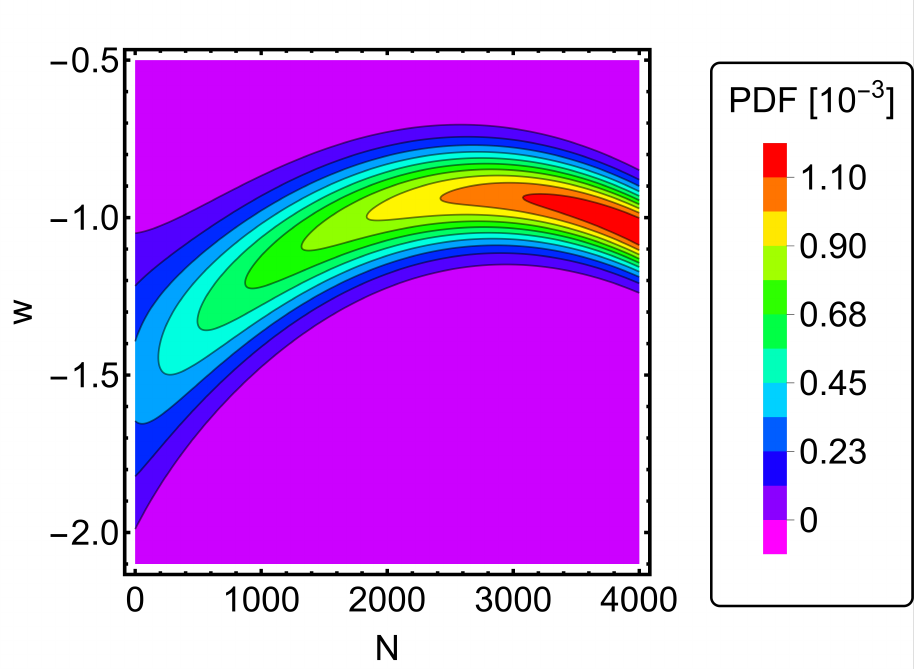}}
\textcolor{blue}{\hspace{0.35cm}}
\subfloat[All parameters of the X-ray correlation are fixed.]{\includegraphics[width=2in]{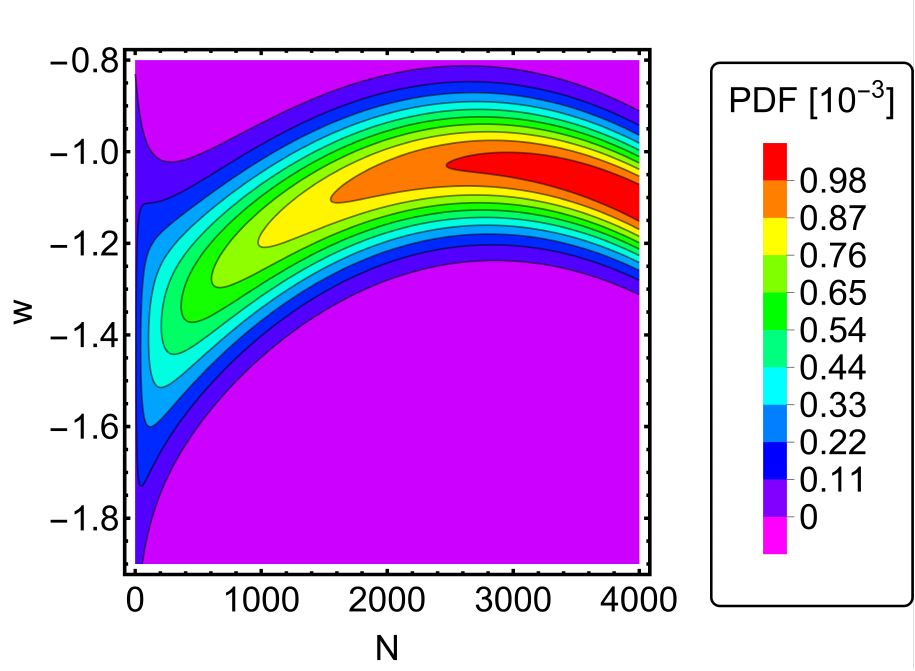}}\\

\subfloat[All parameters of the optical correlation are varied.]{\includegraphics[width=2in]{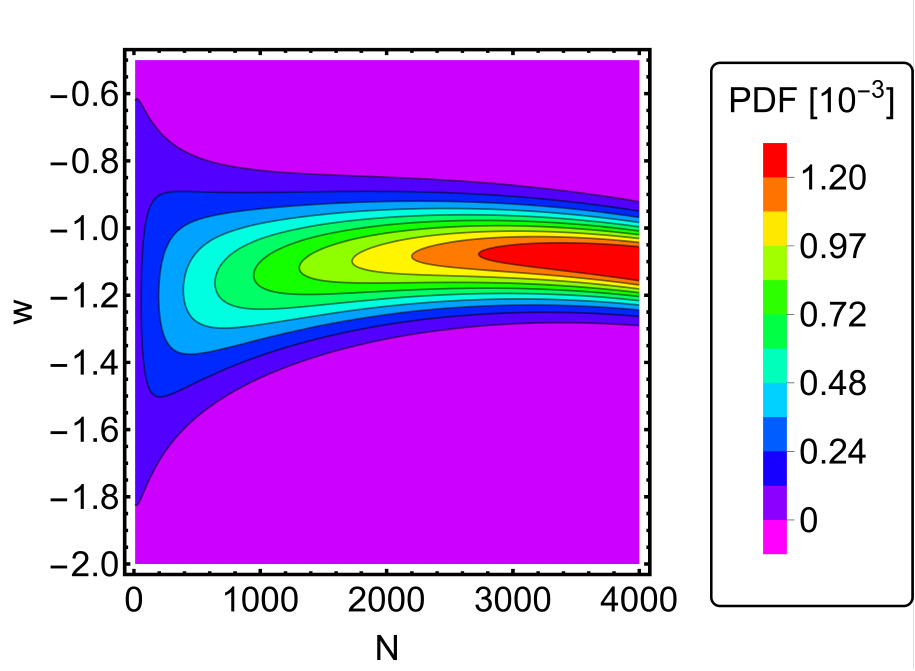}}
\textcolor{blue}{\hspace{0.35cm}}
\subfloat[The normalisation of the optical correlation is fixed.]{\includegraphics[width=2in]{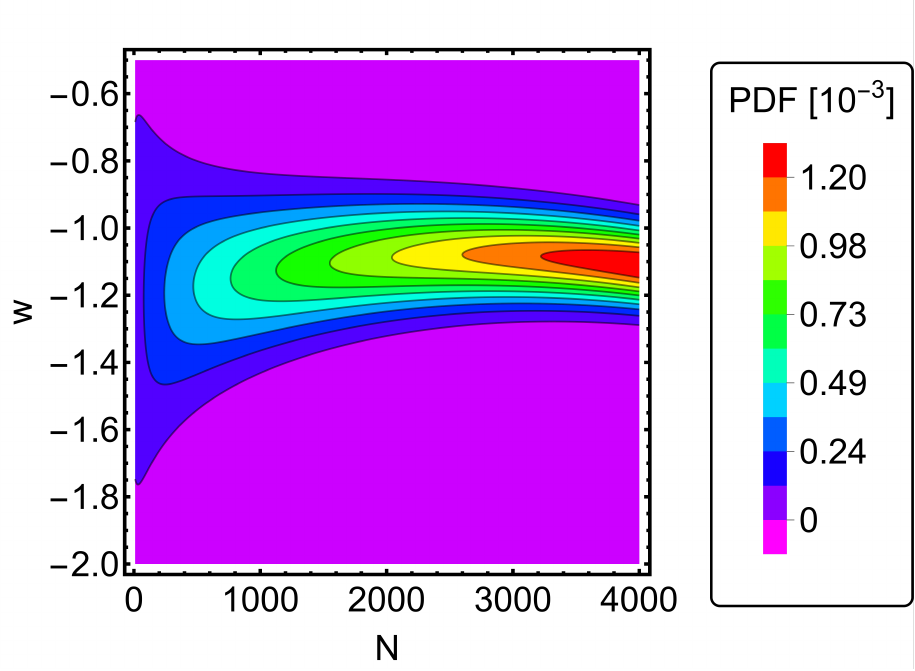}}
\textcolor{blue}{\hspace{0.35cm}}
\subfloat[All parameters of the optical correlation are fixed.]{\includegraphics[width=2in]{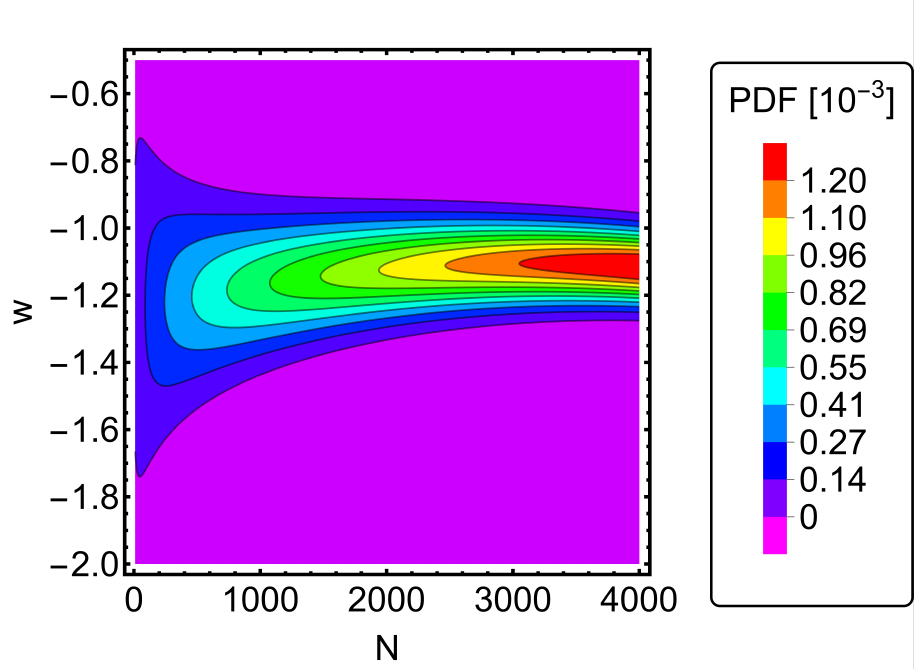}}

\caption{2D X-ray (top row) and optical (bottom row) mock GRBs analysed with simulated Planck data.}
\label{fig:w2DMockCMB}
\end{figure*}

This simulated setup allows us to reach the precision of DESI on the $w$ parameter with 1631 (1358) X-ray (optical) GRBs analysed with the correlation parameters left free to vary. When the normalisation of the correlation is fixed, this number reduces to 1473 (1320). Additionally, when all parameters of the correlation are fixed, the resulting number is 666 (1052). The accuracy has improved significantly compared to the results obtained with real Planck data. Figure~\ref{fig:w2DMockCMB} demonstrates that the optical sample has more stable precision than the X-ray sample.

\subsection{The precision of $w_{0}w_{a}$CDM model with GRBs and Planck}

\begin{figure*}[ht!]
\centering
\subfloat[Constrains obtained with X-ray sample.]{\includegraphics[width=0.49\textwidth]{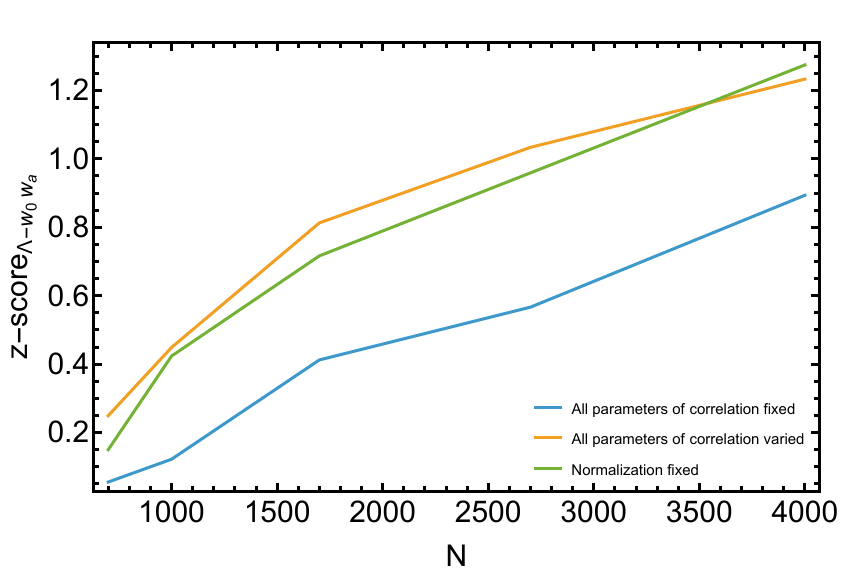}}
\subfloat[Constrains obtained with optical sample.]{\includegraphics[width=0.49\textwidth]{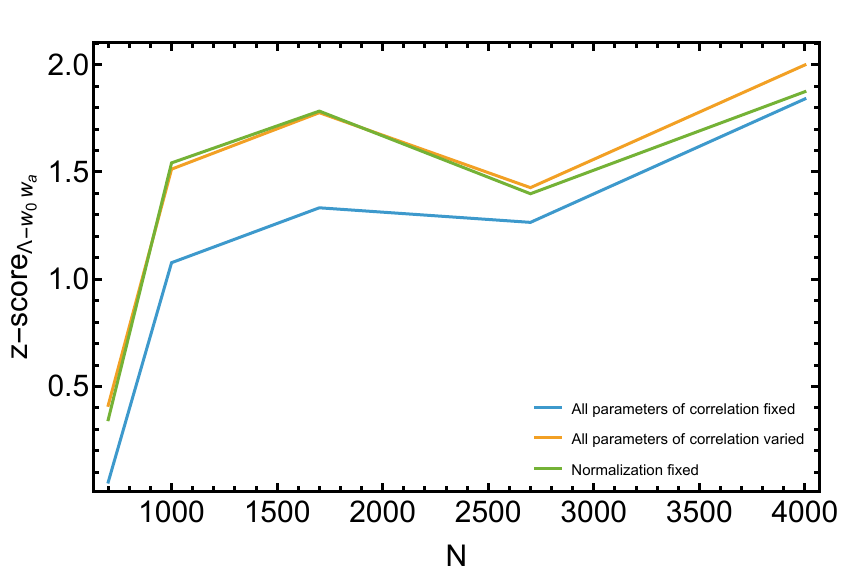}}
\caption{The z-score between $\Lambda$CDM and $w_0w_a$CDM models computed with Equation~\ref{eq:zmodels}. 2D mock GRBs are analysed with mock Planck data.}
\label{fig:w0wa}
\end{figure*}

Our results indicate that, for GRB samples with $\leq 4000$ sources, it is not possible to constrain $w_a$ when both $w_0$ and $w_a$ are allowed to vary. When we fixed $w_a=-1.09$, we could not confirm $w_0>-1$ at the $3\sigma$ level with $\leq 4000$ GRBs. However, when we fixed $w_0=-0.667$, we were able to confirm a non-zero value of $w_a$. 

In the case of GRBs analysed with mock Planck, when all parameters of the correlation are left free to vary, one needs 2340 (1594) sources in the X-ray (optical) case. When the normalisation of the correlation is fixed, one needs 2399 (1513) X-ray (optical) sources. Finally, when all the parameters of the correlation are fixed, one needs 1793 (1329) X-ray (optical) sources.

Nonetheless, for samples of 4000 GRBs in both X-ray and optical cases, we obtain contours closed outside the $\Lambda$CDM point $w_0=-1$, $w_a=0$ within $3\sigma$. While this is encouraging, it does not imply that the $w_0w_a$CDM model is favoured over $\Lambda$CDM at the 3$\sigma$ level. To compare two models in such a scenario, one would need to apply Wilks' theorem~\citep{Wilks1938}. However, this is not applicable here, because our posteriors indicate only an upper limit on $w_a$. In such a case, to compare the two models, we introduce a pivot parameter $w_p$, which decorrelates the $w_0$ and $w_a$ parameters:
\begin{equation}
    w_p = w_0 + (1-a_p)w_a,
\end{equation}
where $a_p$ is the pivot scale factor, chosen in a way which makes $w_p$ and $w_a$ uncorrelated and minimizes the uncertainty on $w_p$: $a_p = 1+ \frac{\text{Cov}(w_0,w_a)}{\text{Var}(w_a)}$~\citep{Hu2004}. The point $w_p=-1$ corresponds to the $\Lambda$CDM model. Therefore, we can estimate the probability of our best-fit being in disagreement with the $\Lambda$CDM model by computing a simple z-score:
\begin{equation}
    z=\frac{|w_{\rm p,\, best-fit}-w_{\rm p,\, \Lambda CDM}|}{\sigma_{w_p}}=\frac{|w_{\rm p,\, best-fit}+1|}{\sigma_{w_p}},
    \label{eq:zmodels}
\end{equation}
where $\sigma_{w_p}$ is the uncertainty on the $w_p$ parameter.

In all our simulations, we obtained that the $z$ increases when we increase the number of GRBs, as one would expect (see Figure~\ref{fig:w0wa}). With 4000 sources, we obtained a maximum of $z\approx1.3$ (in the X-ray case with fixed normalisation) and $z\approx2$ (in the optical case with all parameters of the correlation varied). %This is a comparable result to the one obtained with DESI+BAO data~\citep{DESI2025}. Still, the z-score grows quite quickly with the number of sources. Therefore, the combination of the X-ray and optical data, as well as a potential combination with other GRB correlations like the~\citep{Amati+02} correlation, could enable $3\sigma$ constraints within the next two decades.
This is a comparable result to the one reported from DESI+BAO data~\citep{DESI2025}. Moreover, the z-score increases steadily with the number of GRBs. This suggests that combining the X-ray and optical samples, and potentially including additional GRB correlations such as the Amati relation~\citep{Amati+02}, could further improve the constraining power in the future.

%In summary, although we can witness the constraining power of GRBs when we compare them with DESI alone, these are still marginal compared to the combination of DESI BAO + CMB + SNe Ia altogether. 
In summary, although GRBs impact remains limited compared to the full DESI BAO + CMB + SNe Ia combination, they already show constraining power when compared with DESI alone. Hence, GRBs provide an independent high-redshift probe of the expansion history, and therefore have significant value as a complementary test of cosmological models. Moreover, this analysis provides a reliable and conservative estimate of the current and near future expectations on GRB cosmology.

\section{Future GRB observations and development}
\label{future}

\subsection{The current state-of-the-art in Machine Learning}

Recent advances in ML techniques have opened new possibilities for improving the precision of GRB observables used in cosmological analyses. In particular, ML algorithms can reconstruct GRB light curves, identify plateau phases more robustly, and estimate key parameters such as the rest-frame plateau time and luminosity with reduced uncertainties. Several studies have demonstrated that ML-based reconstructions of GRB light curves can decrease the statistical uncertainties on the fitted parameters by improving the sampling of poorly observed portions of the light curve and by mitigating instrumental noise effects. In addition, ML approaches are increasingly used to estimate redshifts from prompt and afterglow properties, enabling the inclusion of GRBs without spectroscopic redshift measurements. The combination of improved parameter estimation and redshift inference can significantly enlarge the usable GRB sample and reduce the uncertainties associated with the Dainotti relation. In the context of the present work, we model this improvement conservatively by assuming a reduction in the observational uncertainties of the light-curve parameters by a factor of two, which reflects the expected gain achievable with modern ML-based light-curve reconstruction techniques.

Another important method which is able to increase the sample size is the transfer learning tool in which knowledge gained from a source domain, $\mathcal{D}_S$, is reused to improve performance in a related target domain, $\mathcal{D}_T$, to enhance the predictive function by transferring learned representations, model parameters, or feature embeddings from $\mathcal{D}_S$, provided that the two domains share an underlying structure.

In the case of GRBs, optical and X-ray afterglows originate from the same external shock synchrotron emission and therefore exhibit correlated temporal and luminosity evolution. The existence of analogous 2D Dainotti relations in both bands suggests that their latent feature spaces are closely connected. This physical consistency motivates the application of transfer learning between the optical and X-ray samples.

Operationally, we consider (Dainotti, Bogdan et al.~2026, in preparation) transferring a model trained on GRBs with known redshift in one band (e.g., X-rays) and fine-tuning it on the other band (e.g., optical). The shared feature-extraction layers capture the plateau temporal--luminosity behavior, while the final regression layers are adapted to the target dataset. In this way, redshifts can be inferred for GRBs lacking spectroscopic measurements in a single band by leveraging information learned from the other band.
The main advantage of this approach is an effective enlargement of the cosmologically usable sample. Since ML methods will double the fraction of GRBs with reliable inferred redshift and transfer learning allows cross-band generalization, the effective sample size will increase by up to a factor of $\sim 5.4$ compared to the current optical sample only\textcolor{blue}{.} Since parameter uncertainties approximately scale as $\sigma \propto N^{-1/2}$, this corresponds to nearly halving the time required to reach a given precision on $w$.

Therefore, transfer learning provides a statistically efficient and physically motivated strategy to combine optical and X-ray GRB datasets into a unified framework for high-redshift cosmology.

Optical and X-ray samples have very similar constraining power. Thus, we estimate the number of years needed to achieve our goals by combining the X-ray and optical samples. %We discuss here the best-case scenarios, while Table~\ref{tab:results_rotated} presents all cases. If we vary all parameters simultaneously, we can reach Planck precision in 5 years using GRBs alone for X-ray data (corresponding to a reduction of 6 years). If we consider instead the smallest number of GRBs needed, we can reach such precision in 4 years for the optical data.
We discuss here the best-case scenarios, while Table~\ref{tab:results_rotated} presents all cases. If all parameters are varied simultaneously, Planck precision can be reached in 5 years using X-ray GRBs alone, corresponding to a gain of 6 years. If, instead, we consider the case requiring the smallest number of GRBs, the same precision can be reached in 4 years using the optical data.
If we fix the parameters of the correlation, which is equivalent to calibrating the relation on other probes, we can reach the same precision as Planck this year, both for optical and X-rays.
If we aim to reach the precision of DESI with GRBs and Planck real data, this will be reached in 2032 with X-ray data, both varying all parameters or calibrating them. If we calibrate our GRBs with mock Planck data, we can reach this precision in 2030.

\begin{figure}[ht!]
    \centering
    \includegraphics[width=0.49\linewidth]{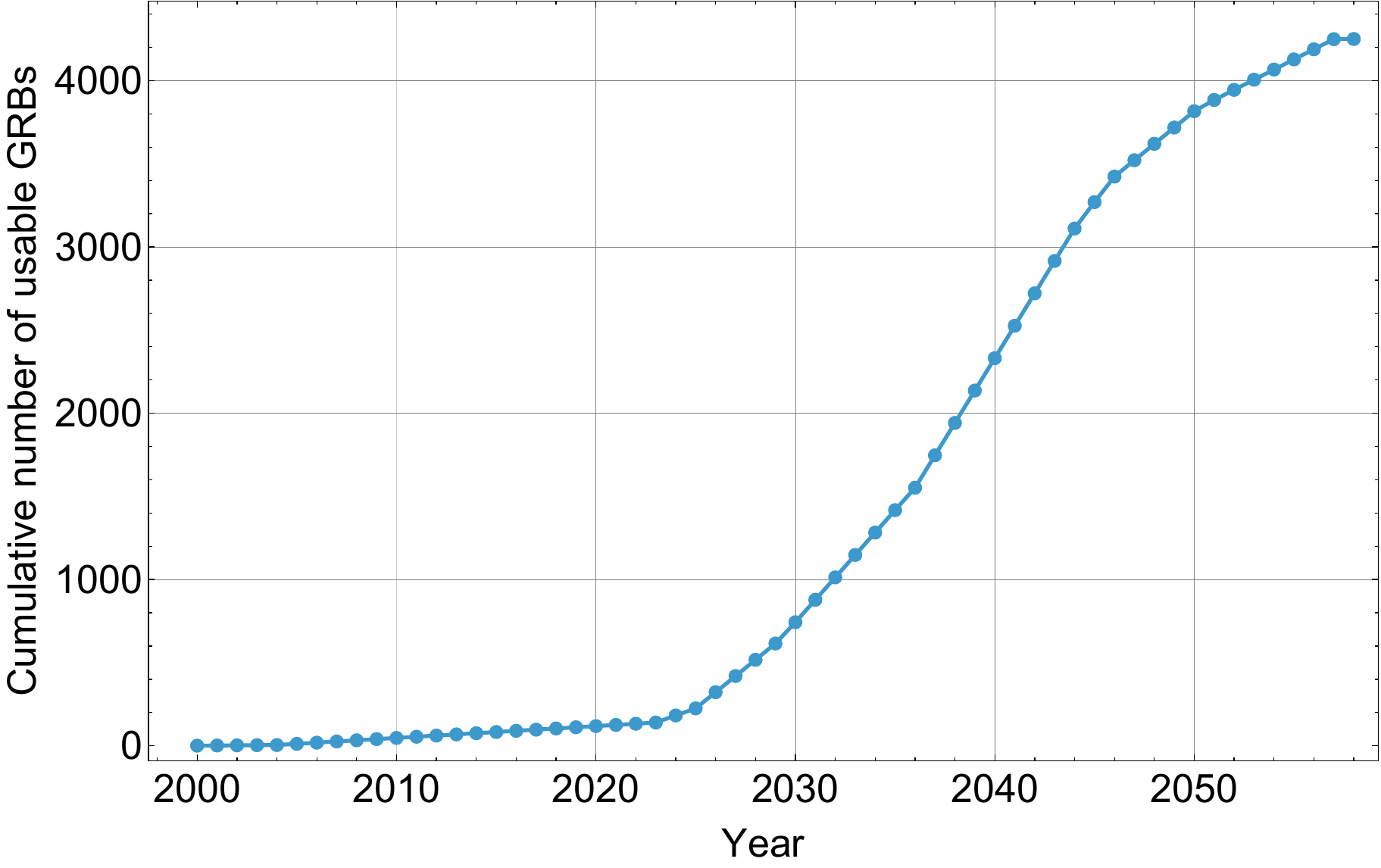}
    \caption{Estimated total number of X-ray and optical GRB afterglow detections by current and possible future instruments as a function of years.}
    \label{fig:future_transfer_learning}
\end{figure}

\subsection{Future observations}

Currently, the majority of GRB X-ray afterglows are observed by the Swift satellite~\citep{2004ApJ...611.1005G}. However, the recent start of the SVOM~\citep{2016arXiv161006892W} and Einstein Probe~\citep{EP2025SCPMA..6839501Y} missions and the development of planned missions like Hiz-GUNDAM or THESEUS will ensure a much larger number of observations in the future. Our interest is in when we will have a sufficient number of GRBs to constrain the $w$ parameter with a precision comparable to Planck and DESI.
Therefore, in Table~\ref{tab:future}, we include the estimated rate of detections of X-ray afterglows, the planned first light of the instrument, and the expected date of the end of each mission, based on the life expectancy of the current instruments.

\begin{table}[ht!]
\centering
\begin{tabular}{lcccc}
\hline
\multicolumn{5}{c}{\textbf{X-ray}} \\
\hline
\hline
\textbf{Mission} & \textbf{Rate of afterglows [yr$^{-1}$]} & \textbf{Start} & \textbf{End} & Reference \\
\hline
Swift                    & 72 & 2005 & 2045 & Computed here, ~\cite{2004ApJ...611.1005G}\\
SVOM                     & 130   & 2024 & 2044 & \href{https://fsc.svom.org/ifsc-tools/grb-public}{SVOM catalogue}\\
Einstein Probe & 370   & 2024 & 2044 & ~\citep{EP2025SCPMA..6839501Y}\\
eXTP                     & 300   & 2030 & 2050 &~\cite{eXTP2025}\\
Hiz-GUNDAM               & 125   & 2030 & 2050 & \href{https://nagataki-lab.riken.jp/workshop/GRB2015/0901/FINAL-RIKEN-HiZ-GUNDAM-YONETOKU.pdf}{RIKEN website}\\
STROBE-X\footnote{This mission was not accepted for further stages of development, however, we still keep it to demonstrate its capabilities}                 & 100   & 2031 & 2051 & \href{https://strobe-x.org/STROBE_X_Study_Report-FINAL.pdf}{Official STROBE-X website}\\
THESEUS                  & 700   & 2037 & 2057 &~\cite{2018AdSpR..62..191A} \\
\hline
\hline
\multicolumn{5}{c}{\textbf{Optical}} \\
\hline
\textbf{Facility / Instrument} & \textbf{Rate [yr$^{-1}$]} & \textbf{Start} & \textbf{End} & Reference \\
\hline
Swift UVOT                 & 27.4  & 2005 & 2045 & Calculated in this work \\
SVOM VT                    & 49  & 2024 & 2044 & 38\% of X-ray \\
Ground Robotic Telescopes  & 18 & 2000 & - & Calculated in this work \\
LSST / Rubin Observatory   & 50  & 2026 & 2046 &~\cite{LSSTscienceBook}\\
LSST Orphan afterglows   & 1000  & 2026 & 2046 &~\cite{LSSTscienceBook}\\
%ELT-class Telescopes       & 40?  & 2028 & 2050 & \\
Subaru-Ultimate & 8 & 2028 & 2048 & Calculated in this work \\
Hiz-GUNDAM (optical)       & 47.5  & 2030 & 2050 & 38\% of X-ray \\
THESEUS (optical)            & 266  & 2037 & 2057 & 38\% of X-ray \\
\hline
\end{tabular}
\caption{Estimated GRB/X-ray mission rates and operational windows (optimistic extensions are included).}
\label{tab:future}
\end{table}

\subsection{X-ray missions}

The rate of Swift XRT detections was calculated based on the past $\sim20$ years of operation, resulting in $\sim 72$ events per year~\citep{DainottiVia2022MNRAS.514.1828D}. We assume that in the near future its orbit will be corrected\footnote{\url{https://www.nasa.gov/news-release/nasa-awards-company-to-attempt-swift-spacecraft-orbit-boost/}} and its operational time extended until 2045. Through 2025, SVOM observed 130 sources\footnote{\url{https://fsc.svom.org/ifsc-tools/grb-public}}, which is our assumed rate of observations per year. We assume that most of the detected sources will be observed by the X-ray instruments onboard. Its original mission end date is set to 2027; however, given its successful operation, we expect it will be extended for a total of 20 years, as is usually the case with X-ray missions. We also use this period for all the other missions. 
There are very few resources available regarding the Einstein Probe sensitivity for the total GRB population. However, a recent study gives a rough estimate of $\sim 370$ events per year~\citep{EP2025SCPMA..6839501Y,2004ApJ...611.1005G}.  Its sensitivity to high-$z$ GRBs is expected to be 3 times higher than that of Swift~\citep{EinsteinProbe2025}. Thus, this mission might be very valuable for cosmological studies. One can find information on the RIKEN website that Hiz-GUNDAM will observe $\sim 100$ ``normal'' GRBs in X-rays per year\footnote{\url{https://nagataki-lab.riken.jp/workshop/GRB2015/0901/FINAL-RIKEN-HiZ-GUNDAM-YONETOKU.pdf}}. 
If STROBE-X were launched, it would observe $\sim 100$ ``canonical'' GRBs in X-rays per year\footnote{More information available at the mission's official website: \url{https://strobe-x.org/STROBE_X_Study_Report-FINAL.pdf}}. eXTP is a polarization-focused mission; however, it will observe between 173 and 328 plateaus per year~\citep{eXTP2025}, thus proving helpful for our estimates. We assume an optimistic value of 300 plateau detections per year. 
It is hypothesized that THESEUS will detect between 300 and 700 GRBs annually~\citep{2018AdSpR..62..191A}. Again, we take an optimistic assumption of 700 sources discovered per year.
The sample of 42 sources used as a parent sample in our analysis constitutes $\sim3.3\%$ of all X-ray afterglows ($\sim 1400$) observed until the end of 2023 (the date of the latest GRB we investigated) with measured redshifts. The total number of GRBs with and without redshifts is similar. However, recent improvements in ML-based redshift inference for GRBs~\citep{2025A&A...698A..92N} may enable us to double the sample size. We also need to consider technological advances in new observatories, which are designed to gather redshift information much more efficiently. Therefore, we expect a much bigger fraction of data with known redshift in the future.
However, here we perform a conservative estimate based on ML, and thus we assume that the fraction of useful GRBs will be double the current one, i.e., $\sim 6.6\%$. To compute the number of cosmologically useful X-ray afterglows, we compute the total number of possible afterglow detections and multiply it by the aforementioned fraction of $6.6\%$. The results of this estimation are presented in the left panel of Figure~\ref{fig:future}.

\subsection{Optical missions}

\begin{figure}[ht!]
    \centering
    \includegraphics[width=0.49\linewidth]{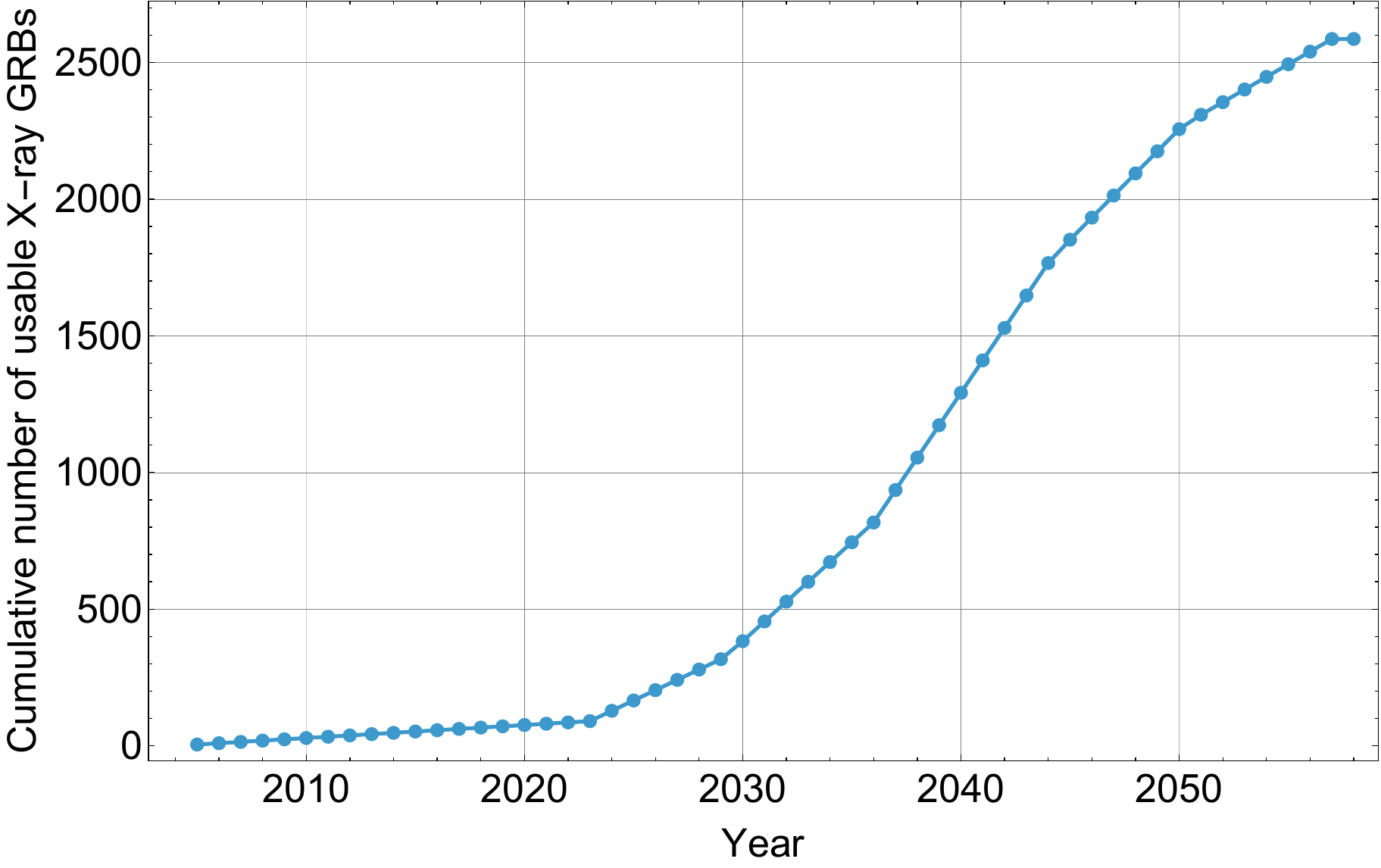}
    \includegraphics[width=0.49\linewidth]{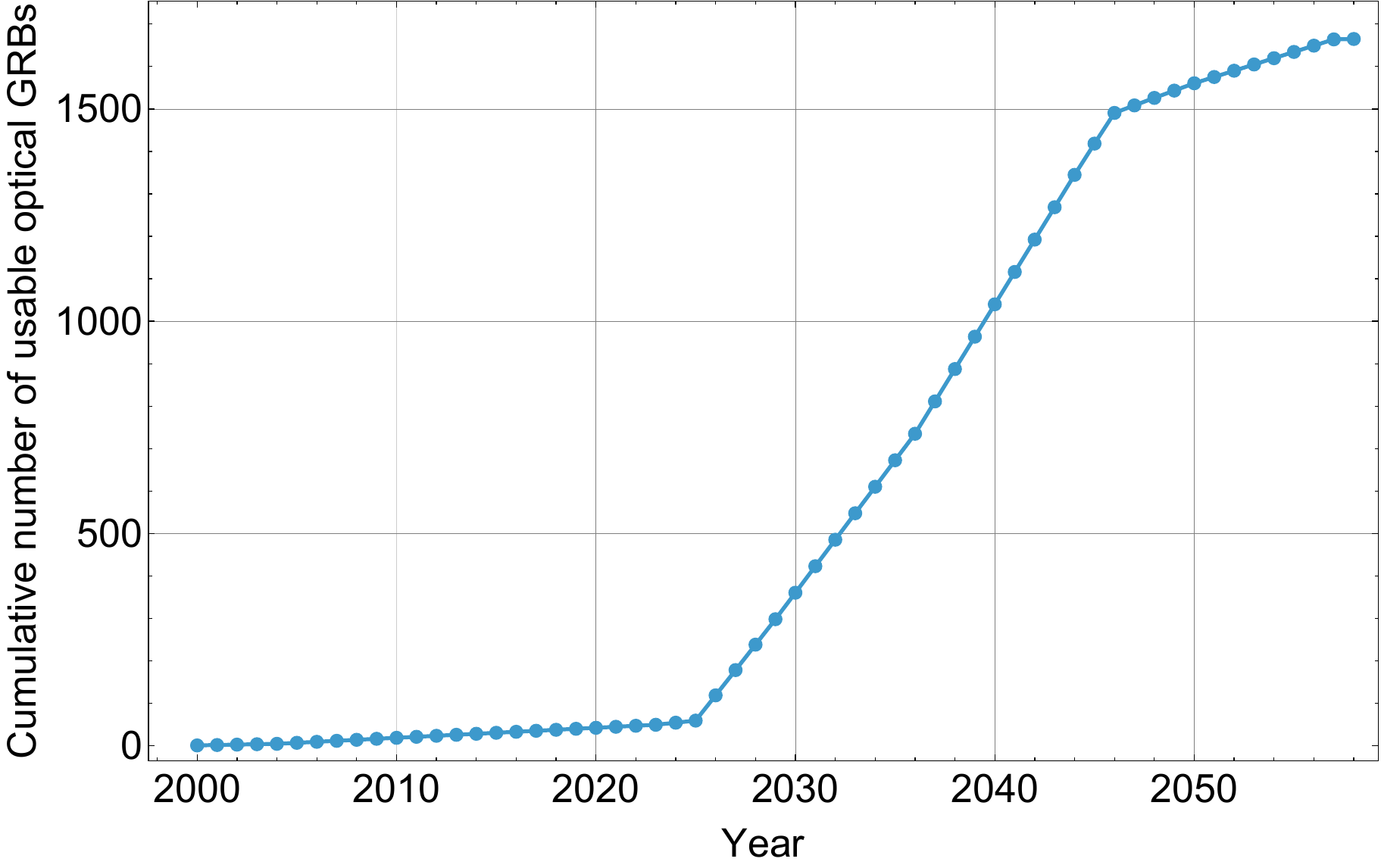}
    \caption{Left: Estimated total number of X-ray GRB afterglow detections by current and possible future instruments as a function of year. Right: The same, but for optical detections.}
    \label{fig:future}
\end{figure}

Around $38\%$ of GRBs detected in X-rays also have an optical detection. Most of those detections come from Swift's UVOT instrument. Therefore, we assume that all X-ray probes will have a similar ratio, except for eXTP and STROBE-X, which will not be equipped with optical telescopes. Additionally, the recently launched Vera C. Rubin Observatory's Large Synoptic Survey Telescope (LSST, \citealt{2023PASP..135j5002H}) is expected to provide $\sim 50$ optical GRBs per year, given its constant monitoring of a large part of the sky. Ground observatories have detected 907 GRBs up to date (04.02.2026); however, 547 overlap with the Swift UVOT detections\footnote{see \url{https://swift.gsfc.nasa.gov/archive/grb_table/stats/}}. Thus, ground-based observatories contributed 360 sources over $\sim 20$ years, resulting in a rate of $\sim 18$ detections per year. Subaru-Ultimate will observe GRBs mostly through follow-up observations with immediate access to $\frac{1}{4}$ of the sky. Its high limiting magnitude in the R-band ($\sim 27$) should enable the detection of emission from the majority of satellite-detected triggers. Given its small field of view, it will be able to observe only sources with well-constrained positions. We assume that $\sim 150$ triggers (almost all triggers of Swift and a quarter of Fermi's) fulfill this criterion. Additionally, we use a conservative estimate that only about 20\% of all triggers will be accepted as targets of opportunity, resulting in about $\sim 8$ possible GRB detections per year.
Our parent sample of 24 sources constitutes $\sim 2.6\%$ of all observed optical afterglows. Again, ML will double this percentage. Thus, we assume that in the future $\sim 5.2\%$ of all optically detected afterglows will be cosmologically useful. As in the case of X-ray data, we computed the cumulative number of observed sources and multiplied it by this factor to obtain the estimated number of high-quality sources. The results of this computation are presented in the right panel of Figure~\ref{fig:future}.

\section{Summary, Discussion and Conclusions}\label{conclusions}

Our results have demonstrated the usefulness of GRBs as cosmological probes that complement traditional distance indicators, particularly by extending the exploration of the expansion history of the Universe to high redshift.
In this work, we have used GRBs within the $w$CDM and $w_0w_a$CDM frameworks, focusing on the constraining power of plateau correlations in both X-ray and optical wavelengths. Using the observed GRB plateau samples as a starting point, we constructed simulated GRB populations and performed cosmological parameter estimation through an MCMC analysis. Our goal is to determine the number of GRBs required to achieve a given level of precision (the one of Planck and DESI) on the dark-energy equation-of-state parameter $w$, both using GRBs alone and in combination with CMB observations.
Our analysis shows that GRBs alone can provide meaningful constraints on cosmological parameters once sufficiently large samples are available, with the precision improving significantly as the sample size increases. In particular, we find that a high-quality GRB sample containing several hundred sources is sufficient to approach the level of precision currently achieved by CMB observations. The exact number depends on the treatment of the parameters of the Dainotti relation and on the assumptions regarding calibration and observational uncertainties.
With the aid of ML, this precision could be obtained even today. Inferring redshift with such statistical methods doubles the current useful GRB sample ~\citep{2025A&A...698A..92N,Dainotti2024ApJ...967L..30D}, which demonstrates the power of this approach. Additionally, the recent redshift classifier will allow us to follow up more GRBs with the support of the community, \citep{Dainotti2025classifier,Sarkar2025}, thus further increasing the available sample in the future. Moreover, we discuss that the state-of-the-art technique, called transfer learning, will further increase the cosmologically useful sample by combining the X-ray and optical data. We refer to Table \ref{tab:results} for all the results.
When combined with CMB data, GRB observations can further improve the constraints on the dark-energy equation of state, demonstrating the complementarity between high-redshift probes and early-Universe measurements.
We can reach the same precision on $w$ as DESI, in 11 years or 6 years if we use transfer learning. If instead we use a mock Planck sample with X-ray data, we can reach the same precision as DESI in 9 years or 4 years if we use transfer learning. 
In this regard, GRBs could provide an important \textit{independent} probe of the Dark Energy equation of state at high redshift, offering a valuable cross-check of possible indications for DDE emerging from other cosmological datasets.
The quoted estimates are conservative, since we will have more follow-up observations, especially at high-redshift, thanks to our high-redshift classifier both in X-rays~\citep{Dainotti2025classifier} and optical~\citep{Sarkar2025}. 

We have also explored the potential of future GRB samples to constrain dynamical dark-energy scenarios within the $w_0w_a$CDM model. Our forecasts indicate that constraining both $w_0$ and $w_a$ simultaneously remains challenging, reflecting both the intrinsic degeneracies of the CPL parametrization and the current limitations in observational precision. Nevertheless, GRBs remain particularly valuable because they extend the cosmological distance ladder to redshifts far beyond those accessible with Type Ia SNe, allowing tests of the expansion history of the Universe in a redshift regime that is still relatively unexplored.
This makes GRBs especially relevant as an independent test of the DDE scenario suggested by recent low-redshift observations~\citep{DESI2025,DES:2025sig,Hoyt:2026fve}.
While current GRB samples remain limited in size, forthcoming space missions and large-scale observational programs are expected to dramatically increase the number of GRBs with well-characterized afterglows. Missions such as SVOM and Einstein Probe, together with future facilities like THESEUS and wide-field optical surveys such as the Vera C. Rubin Observatory (LSST), will significantly improve both the detection rate and the characterization of GRB plateau emissions. In addition, advances in analysis techniques, including machine-learning-based light-curve reconstruction and redshift inference, will help reduce observational uncertainties and expand the number of GRBs usable for cosmological studies.
In addition to the individual correlations discussed above, GRBs exhibit multiple empirical relations that can be jointly exploited for cosmological analyses. 
In particular, most of the GRBs that obey the Dainotti correlation also follow the Amati correlation ~\citep{Lenart2025}. Although these correlations are the most promising, due to the simplicity of measuring the parameters involved, GRBs also follow a variability--luminosity correlation, which might further help in such studies ~\citep{Schaefer2007}. 
If selection biases are properly accounted for in the other correlations as well, their combination will further improve the precision on cosmological parameters.
Given Equation \ref{eq:totaluncertainty}, the combination of three correlations with similar sample sizes should reduce the uncertainty by an additional factor of $\sim \sqrt{3}$. 
Within the broader cosmological context, it is worth noting that current analyses remain broadly consistent with a cosmological constant, while still allowing for the possibility of DDE models. Analyses combining CMB data with large-scale structure and SNe measurements indicate that the equation-of-state parameter is compatible with $\Lambda$CDM within present uncertainties (e.g., \cite{Planck:2019nip}; \cite{DiValentino:2021izs}). In this sense, DDE models remain subdominant descriptions of the current observational landscape, although they continue to be well motivated from a theoretical standpoint and probing these scenarios requires observational tests across a wide redshift range where GRBs provide a unique contribution.
The results described in detail in Sec. \ref{results} show that the constraints obtained in the $w$CDM framework are consistent with the $\Lambda$CDM scenario, as the fiducial value $w=-1$ is recovered within the explored configurations. 
As shown in Table 2, several configurations—particularly within the $w_0w_a$CDM framework—exhibit best-fit values that depart from the $\Lambda$CDM expectations, especially in the parameters $w_0$ and $w_a$. Although these deviations remain within the current uncertainties, they indicate a measurable sensitivity of GRB-based analyses to potential departures from $\Lambda$CDM at high redshift.
Thus, these results provide a quantitative assessment of the sample size and observational requirements needed for GRBs to become competitive cosmological probes, particularly at high redshift.
In this context, GRBs become an increasingly important tool for testing extensions of the standard cosmological model and probing the evolution of Dark Energy across a broad redshift range. The forecasts presented in this work, therefore, provide a quantitative roadmap for the role that GRBs will play in future cosmological studies. In the coming decade, the combination of next-generation GRB missions, wide-field optical surveys, and improved statistical and machine-learning techniques will transform GRBs into one of the few astrophysical probes capable of tracing the expansion history of the Universe from the local Universe to the highest observable redshifts.

\begin{acknowledgments}

AŁL acknowledges that the publication has been supported by a grant from the Faculty of Physics, Astronomy, and Computer Science under the Strategic Programme Excellence Initiative at Jagiellonian 
University. EDV is supported by a Royal Society Dorothy Hodgkin Research Fellowship and she acknowledges the support from NAOJ and the Division of Science at NAOJ for her visit to MGD at NAOJ. NF is grateful to UNAM-DGAPA-PAPIIT for the financial support granted via IN112525.  WG acknowledges support from the National Aeronautics and Space Administration (NASA) under Grant No. 80NSSC24K0898. This article is based upon work from COST Action CA21136, Addressing observational tensions in cosmology with systematics and fundamental physics (CosmoVerse)+ supported by COST (European Cooperation in Science and Technology).
M.G.D. acknowledges the support of the JSPS Grant-in-Aid for Scientific Research (KAKENHI) (A), Grant Number JP25H00675. Numerical computations were in part carried out on Cray XD2000 at the Center for Computational Astrophysics, National Astronomical Observatory of Japan. We acknowledge the IT Services at The University of Sheffield for the provision of services for High Performance Computing. 

\end{acknowledgments}

\bibliography{bibliography}{}
\bibliographystyle{aasjournalv7}

\end{document}